\documentclass[11pt]{article}
\usepackage{latexsym}
\usepackage{amsmath}
\usepackage{amsfonts}
\usepackage{mathrsfs}
\usepackage{dsfont}
\usepackage{verbatim}
\usepackage{hyperref}
\usepackage{tikz}
\usetikzlibrary{matrix,arrows}
\usepackage{subfigure}
\usepackage{mathtools}
\usepackage{simplewick}

 \csname
@addtoreset\endcsname{equation}{section}
\usepackage{amsmath,amssymb}
\usepackage{graphicx}
\usepackage{color}

\begin{document}

\def\be{\begin{equation}}
\def\ee{\end{equation}}
\def\bea{\begin{eqnarray}}
\def\eea{\end{eqnarray}}
\def\nn{\nonumber}
\def\n{n+\delta'}
\def\barn{\bar{n} + \bar{\delta}'}
\def\m{m+\delta''}
\def\barm{\bar{m}+\bar{\delta}''}

\begin{flushright}

\end{flushright}

\vspace{40pt}

\begin{center}

{\Large\sc On scalar propagators of three-dimensional higher-spin black holes}

\vspace{50pt}

{\sc Hai Siong Tan}

\vspace{15pt}
{\sl\small
Division of Physics and Applied Physics,
School of Physical and Mathematical Sciences, \\
Nanyang Technological University,\\
21 Nanyang Link, Singapore 637371, \\

\vspace{10pt}
{\it HS.Tan@ntu.edu.sg}}

\vspace{70pt} {\sc\large Abstract}\end{center}

We explore some aspects of three-dimensional higher-spin holography by studying scalar fluctuations 
in the background of higher-spin black holes. We furnish an independent derivation of the bulk-boundary propagator by purely invoking a well-known infinite dimensional matrix representation of $hs[\lambda]$ algebra related to its construction as a quotient of the universal enveloping algebra of $sl(2)$, thus evading the need in previous literature to perform an analytic continuation from some integer to $\lambda$. The propagator and the boundary two-point functions are derived for black hole solutions in $hs[\lambda]\times hs[\lambda]$ Chern-Simons theory with spin-3 and spin-4 charges up to second-order in the potentials. We match them with three- and four-point torus correlation functions of the putative dual conformal field theory which has $\mathcal{W}_\infty [\lambda]$ symmetry and is deformed by higher-spin currents. 
\newpage

\tableofcontents
\section{Introduction}
\label{sec:Intro}

Higher-spin gravity in three dimensions has furnished a fruitful arena for exploring holography-based techniques which are useful for understanding quantum gravity. In this case, the holographic duality lies between a three-dimensional bulk quantum gravity theory (of the Vasiliev type) \cite{Vas} and two-dimensional CFTs with higher spin symmetry. In this paper, we work in the context of a well-known proposal by Gaberdiel and Gopakumar which relates three-dimensional higher-spin gravity to $\mathcal{W}_N$ minimal models which admit coset descriptions \cite{Ga1,Ga2,Ga3}. 

As developed in recent literature, such a higher-spin $AdS_3 / CFT_2$ has two main variants which are defined via two distinct manners of adopting the large central charge limit. The first one, as indicated in the original seminal work, is known as the 't Hooft limit where $N$ and level $k$ are infinite with $\lambda = N/(N+k)$ being fixed. An unresolved subtlety is that there are states in the dual CFT which have conformal weights $\sim 1/N$ that do not decouple and which, as far as we know, have not been matched to sensible objects in the bulk. The second variant as proposed in \cite{Ga3,Perl}
involves keeping $N$ fixed while performing an analytic continuation $\lambda \rightarrow -N$. In this limit, the spectrum matches on both sides of the duality but the CFT is in a non-unitary regime, and it remains to see whether quantum effects in the bulk could be consistently understood even in the face of non-unitarity beyond such a semi-classical limit. 

In exploring the $AdS/CFT$ dictionary, despite the above puzzles, there has been concrete and interesting progress made in comparing the bulk and boundary descriptions such as the emergence of $\mathcal{W}$ algebras in asymptotic symmetries \cite{Hen,Camp} and in particular for the purpose of this paper, the matching of correlation functions \cite{3point,4point}. Correlators involving purely higher spin currents were checked to match, a result which reflects the presence of higher spin symmetry on either side of the holographic duality. Three-point functions of the form $\mathcal{O} \mathcal{O} J^{(s)}$ (where $\mathcal{O}$ refers to the operator dual to the bulk scalar and $J^{(s)}$ the spin-$s$ current)
were shown in \cite{3point} 
to match and a class of four-point functions in 
\cite{4point}
which involve two scalar operators and two conical defect operators were also remarkably shown to be precisely identified on both sides. Another class of examples which is of direct relevance to our paper concerns the two-point function of operators belonging to the dual thermal CFT in the presence of a higher-spin deformation. This has been shown in \cite{ProbeCFT} to match to the boundary two-point function in a bulk containing a black hole carrying a spin-3 charge \cite{KG}. Indeed, the bulk gravity theory contains higher-spin generalizations of the BTZ black holes defined rather abstractly by holonomy conditions along the contractible cycle in a solid torus bulk topology \cite{KG,BHreview}. In ordinary three-dimensional gravity, the holonomy condition admits a natural geometric interpretation related to picking a quotient of the underlying thermal $AdS_3$, but in the higher-spin narrative, one may need to extend the notion of spacetime geometry to gauge-invariance of the Vasiliev theory (see for example \cite{Castro} for a highly interesting
recent development). 

Our work here builds on the seminal results in \cite{Probe} and \cite{ProbeCFT} which we now summarize. In \cite{Probe}, the authors studied scalar fluctuations in the background of a spin-3 black hole. In the bulk theory, the scalar field is part of a tower of many scalar fields, all of them auxiliary in nature except for the physical one which can be picked out by a trace operation. The interaction terms between the scalar fields and the gauge field are specified in Vasiliev theory but to use them to write down the set of coupled equations for the physical scalar field could be a non-trivial task. The work in \cite{Probe} took the linearized equation as the starting point, and solved for the bulk-boundary propagator with the known spin-3 black hole Chern-Simons connection. This was worked out for the specific value of $\lambda = \frac{1}{2}$ for the technical reason that the higher-spin algebra in this case can be conveniently parametrized by a set of harmonic oscillators.  Further taking the bulk point to a point at asymptotic infinity ($\rho \rightarrow \infty$), the boundary two-point function was then shown to agree with the thermal correlator for dual CFT operators, at linear order of the spin-3 chemical potential. 

In \cite{ProbeCFT}, the scalar master field equation was solved to first order in the potential for arbitrary $\lambda$ and in the infinite $\rho$ limit,  to second order for some cases of $\lambda = -N$. The caveat is that the partial differential equation for the scalar perturbation turns out to be rather cumbersome and without any hindsight, it may not be clear how does one go about seeking an explicit solution. In \cite{ProbeCFT}, the CFT calculation was smoothly performed to yield a prediction for the boundary two-point function. This was performed to second-order. Presumably, the first-order result could be used to guess a good ansatz for the bulk-boundary propagator for arbitrary $\lambda$ but the second-order correction to the bulk-boundary propagator was not derived. In the infinite $\rho$ limit, for $\lambda = -N$ (and further extended in \cite{resum} for arbitrary $\lambda$ by an analytic continuation of the integer), one can resort to elementary finite $N$ matrix operations and thus some second-order corrections were explicitly presented and nicely verified to match with the dual CFT's computation. A main result of this paper is the derivation of the bulk-boundary propagator to second-order which can be checked to match precisely with the dual CFT's correlator at the boundary. 

As explained in \cite{ProbeCFT}, the dual CFT quantity is the torus two-point function of a scalar field residing in the large $c$ spectrum of the minimal models with the free action being deformed by a holomorphic higher-spin operator. At $m^{\text{th}}$ order in perturbation theory (of the higher-spin potential), holographic duality leads one to match the integrated correlation functions of $\phi \bar{\phi}$ with $m$ spin-3 fields over $m$ copies of the torus. Working in the high temperature regime implies that we can invoke a modular transformation to obtain the correlators by working in the vacuum sector.  Matching
of the first-order correction relates to conformal symmetry on both sides of the duality and normalization of the higher-spin charges, whereas to second-order the matching is sensitive to the form of the classical $\mathcal{W}_\infty [\lambda]$ algebra and involves computation of some four-point functions involving higher-spin fields and the scalar field.

In this paper, we will furnish an independent derivation of the bulk-boundary propagator directly in the context of the $hs[\lambda]$ algebra that governs the bulk gravity theory, without resorting to any $\lambda = -N$ analytic continuation. We do so by using an infinite-dimensional matrix representation of $hs[\lambda]$ that follows naturally from an infinite-dimensional matrix representation of $sl(2)$ for arbitrary spin once we allude to the universal enveloping algebra construction of $hs[\lambda]$ viewed as a subspace of the quotient $U(sl(2))/\langle C_2 - \frac{1}{4}(\lambda^2-1) \rangle$ where $C_2$ is the quadratic $sl(2)$ casimir. The usefulness of this representation of $hs[\lambda]$ was first concretely mentioned and explored in \cite{4point} where it was invoked to compute pure $AdS_3$ scalar propagators and those of chiral deformation backgrounds. We will apply this method for the spin-3 black hole, giving a bona-fide $hs[\lambda]$ calculation to obtain the bulk-boundary propagator, up to second-order in the higher-spin chemical potential. From it, we can take a simple limit to get the boundary two-point function and thus we obtain the correlator at second-order at arbitrary $\lambda$ which we verify to be identical to the CFT calculation presented in \cite{ProbeCFT}. 

We also apply this method to a black hole with spin-4 charge. This solution was first explicitly constructed in \cite{Spin4} (see \cite{Tan} for some discussion of a spin-4 black hole in $sl(4)\times sl(4)$ Chern-Simons theory). 
We compute its  bulk-boundary propagator to first order in the higher-spin chemical potential, and demontrate that it matches with the CFT result at the boundary. Unfortunately, there is a proliferation of terms in the bulk calculation of the second-order correction to the bulk-boundary propagator which we are unable to manage efficiently nor simplify to obtain the boundary two-point function, although for future purposes, we express it in a closed form in terms of elementary integrals in Appendix \ref{sec:AppA}. On the CFT side, we generalize the calculation in \cite{ProbeCFT} to the case where the dual CFT with $\mathcal{W}_{\infty} [\lambda]$ is deformed by a holomorphic spin-4 operator. We match the correlators on both sides to first-order in the higher-spin potential and present a calculation of the two-point function up to second-order in Appendix \ref{sec:AppB}.

With regards to future work along these lines, we hope that our work has also shed light on the computational feasibility for similar computations (such as those recently discussed in \cite{GenHS}) with arbitrary $\lambda$ without relying on the analytic continuation $\lambda \rightarrow -N$. This is relevant whenever one considers the 't Hooft limit of the higher-spin holographic duality. The plan of our paper reads as follows. We begin by reviewing some basic aspects of higher-spin black holes, bulk-boundary propagators, scalar master field equation in Vasiliev gravity, the basic techniques which we employ to compute the propagator and how we relate these bulk observables to boundary correlators in the dual CFT. In Section \ref{sec:Prop}, we proceed to compute the bulk-boundary propagators up to second-order in higher-spin potentials for the spin-3 black hole, and up to first order for the spin-4 black hole. In Section \ref{sec:Boundary}, we present the dual CFT computations to match the scalar
correlator and finally in Section \ref{sec:Discussion}, we summarize our findings. The Appendices \ref{sec:AppA} and \ref{sec:AppB} collect some technical results related to the second-order corrections to the scalar propagator for the spin-4 black hole in both the bulk and boundary theories.

\section{Some Preliminaries}
\label{sec:Prelim}
\subsection{Higher-spin black holes and the scalar master field equation}
We begin by briefly discussing the notion of higher-spin black holes (please see \cite{BHreview} for a more extensive review). In the absence of the master scalar field, the massless and topological sector
of the three-dimensional Vasiliev gravity theory is described by Chern-Simons theory based on the product of two copies of the infinite-dimensional
Lie algebra $hs[\lambda]$ which is equipped with an associative lone star product which we 
denote as $\star$. Let $A,\bar{A}$ denote two independent elements of $hs[\lambda]$. The background 3-manifold is a solid torus homeomorphic to the Euclidean BTZ black hole and we
adopt coordinates $\rho \in (-\infty,\infty), z \sim z+2\pi \sim z+2\pi \tau, \bar{z} \sim \bar{z} + 2\pi \sim \bar{z} + 2\pi \bar{\tau}$, such that $z,\bar{z}$ parametrize the boundary torus. 
We can pick a gauge such that a flat connection for the black hole reads 
\be
\label{flatconnection}
A(\rho, z, \bar{z}) = b^{-1}ab + b^{-1}db, \bar{A}(\rho, z,\bar{z}) = b\bar{a} b^{-1} + bdb^{-1}
\ee
where $b=e^{\rho V^2_0}$, with $V^2_0$ being the Cartan element of an $sl(2)$ subalgebra of $hs[\lambda]$, and the connections $(a,\bar{a})$
being constant with vanishing $a_\rho, \bar{a}_\rho$. With this ansatz, the flat connection condition reduces to $[a_z, a_{\bar{z}}] = [\bar{a}_z, \bar{a}_{\bar{z}}] = 0$. 
By demanding the Wilson loop wrapping the Euclidean temporal circle to be gauge-conjugate to that of the BTZ, one can derive the 
$hs[\lambda] \times hs[\lambda]$-valued
Chern-Simons connections for any higher-spin black hole by solving the eigenvalue constraint equations
\be
\label{eigenvalueConstraint}
\text{Tr}(\omega^n) = \text{Tr}( \omega^n_{BTZ}  ),\qquad \omega = 2\pi ( \tau A_z + \bar{\tau} A_{\bar{z}} ), \qquad n \in \mathbb{Z}.
\ee
In \cite{Spin4} and \cite{KrausPartition}, the connection components were solved via a perturbation series in the higher-spin chemical potentials after
imposing an integrability condition related to the thermodynamics of the black hole solutions. At $\lambda = 0,1$ where the putative CFT's $\mathcal{W}_\infty [\lambda]$
admits a free-field realization, the thermal partition functions can be matched on both sides of the holographic duality (see for example \cite{Boer,Perez} for a deeper discussion on 
higher-spin black hole thermodynamics, etc.). 

We now turn to discussing scalar bulk-boundary propagators. A salient feature of higher-spin gravity of Vasiliev type is a consistent coupling of matter fields to gravity. Let a master field - containing both the physical and auxiliary scalar fields be denoted by $C$. This is a
spacetime scalar transforming in a twisted adjoint representation of the algebra $hs[\lambda]$ following the equation 
\be
\label{MasterKG}
dC + A \star C - C \star \bar{A} = 0
\ee
at the linearized level. The physical scalar field $\Phi$ can be obtained by a trace operation and we can write $\Phi = \text{Tr}(C)$. (We will discuss the trace in greater detail shortly. )
Thus, given some flat connection as the background, in principle, one should decouple the various auxiliary fields and derive a generalized Klein-Gordan equation for $\Phi$ - as was done 
very nicely in \cite{ProbeCFT} and \cite{OtherBH2}. As first explained carefully in \cite{ProbeCFT}, we can use the gauge covariance of various fields to arrive at a convenient ansatz for $C$. Under a
gauge transformation, the fields transform as
\be
A \rightarrow g^{-1} \star \left( d + A \right) \star g,\qquad
\bar{A} \rightarrow \bar{g}^{-1} \star \left( d + \bar{A} \right) \star \bar{g},\qquad
C \rightarrow g^{-1} \star C \star \bar{g}.
\ee
From the gauge where $A=0, dc=0$, $c$ being the master field in this gauge, 
a gauge transformation yields the fields in \eqref{flatconnection}, with
\be
A = g^{-1} dg, \bar{A} = \bar{g}^{-1} d\bar{g}, \qquad g=e^{a_\mu x^\mu}b,
\ee
and the physical scalar field then reads
\be
\label{linearizedscalar}
\Phi (\rho, z,\bar{z}) = e^{\Delta \rho} \text{Tr} \left[ e^{-\Lambda} \star c \star e^{\bar{\Lambda}}  \right],\qquad \Lambda = b^{-1} \star a_\mu x^\mu \star b,
\bar{\Lambda} = b \star \bar{a}_\mu x^\mu \star b^{-1},
\ee
where $\Delta = 1\pm \lambda$ is the conformal dimension of the dual scalar operator. 
It was argued in 
\cite{Probe} that choosing $c$ to be a highest weight state of $hs[\lambda]$ is equivalent to
the higher-spin generalization of the delta-function boundary conditions for the bulk-boundary propagator 
in pure $AdS_3$ gravity. Writing the Euclidean line element as $ds^2 = d\rho^2 + e^{2\rho} dz d\bar{z}$, the ordinary bulk-boundary $AdS_3$ propagator reads 
$$\Phi^{pure\,\, AdS_3}_{\pm} \left( \rho, z, \bar{z} \right) = \pm \frac{\lambda}{\pi} \left(   \frac{e^{-\rho}}{e^{-2\rho} + |z-z'|^2} \right)^{1\pm \lambda}, $$
where the choice of sign $\pm$ indicates the standard and alternate quantization. In the higher-spin case, this is manifest in $\Delta = 1 \pm \lambda$. From now on, we shall stick to the positive sign for definiteness. To relate to the dual CFT, one sends the bulk point to the boundary, taking the
$\rho \rightarrow \pm \infty$ limit. For the higher-spin case, the bulk scalar correlator in the background of a higher-spin black hole was
studied in \cite{Probe} and \cite{ProbeCFT} and the correlator at the boundary was matched precisely with the two-point function of a scalar operator in the CFT with $\mathcal{W}_\infty [\lambda]$ symmetry, and deformed by the corresponding higher-spin operator associated with the black hole's higher-spin charges.
More generally, it was explained in \cite{4point} that solving \eqref{MasterKG} involves choosing a
representation of $hs[\lambda]$ for the gauge fields and master field. In this work, we take the fields
to live in the fundamental representation, noting that in particular for such a setting, the physical scalar field 
and its excitations are dual to the CFT's $(f;0)$ primary and its descendants. 

The aim of this paper is to demonstrate, using higher-spin black hole backgrounds as illustrative examples, that the scalar propagator 
in \eqref{linearizedscalar} can be computed using an infinite-dimensional matrix representation of $hs[\lambda]$. This was first noted in \cite{4point} and we followed one of the future directions stated in that work to realize the computation for higher-spin black holes's propagators. In \cite{ProbeCFT}, the CFT's two-point function was derived up to second-order in the higher-spin potential with the derivation hinging upon 
the form of the classical $\mathcal{W}_\infty [\lambda]$ algebra at large central charge. But the bulk's correlator was not computed for general $\lambda$. Similar to related conclusions in recent literature, the main approach was to compute the bulk's scalar propagator in $sl(N)\times sl(N)$ Chern-Simons theory and then conjecture that one can pass on to $\lambda \rightarrow -N$ via an analytic continuation. In this paper, we will derive the bulk scalar propagator directly in $hs[\lambda]$ theory up to second-order and we found that it matches with the CFT calculation. 

Finally, it is known that in the defining 
representation, the highest weight state $c$ can be viewed as a projector and in an infinite-dimensional matrix representation, it reads $c=\text{Diag} \left(1,0,0,\ldots \right)$, with which the bulk-boundary propagator then reduces to a simple-looking matrix element
\be
\label{PropMatrixelement}
\Phi (\rho, z,\bar{z}) = e^{\Delta \rho} \langle 1 \vert e^{\bar{\Lambda}} e^{-\Lambda} \vert 1\rangle.
\ee
Next, we will review the derivation of \eqref{PropMatrixelement} and discuss how it can be computed explicitly.

\subsection{An infinite-dimensional matrix representation of $hs[\lambda]$}

The higher-spin $hs[\lambda]$ algebra is an infinite-dimensional Lie algebra that admits a simple description 
as a subspace of a particular quotient of the universal enveloping algebra of $U\left(\text{sl}(2) \right)$ which we review below (see for example \cite{SymMin,4point}). The parameter $\lambda$ yields a family of such an algebra of which generators we denote as $V^s_n, \,\,\, s \geq 2, \,\,\, |n|<s.$ In particular, $V^2_{0,\pm 1}$ generate an sl(2) subalgebra
under which $V^s_n$ has spin $s-1$, i.e. 
$$
\left[ V^2_m, V^s_n \right] = \left( -n + m(s-1) \right) V^s_{m+n}.
$$
The physical implication for the higher-spin field theory is that the bulk fields valued in $V^s_n$ have spacetime spin $s$. 
The structure constants can be expressed in terms of hypergeometric functions and 
Pochammer symbols but we will not display them here since we do not require them explicitly. From now, let's denote the sl(2) generators by 
$J_0 \equiv V^2_0, J_{\pm} \equiv V^2_{\pm 1}$. Consider the following quotient of the $U(sl(2))$.
\be
B\left[\mu \right] = U(sl(2))/\langle C_2 - \mu \mathds{1} \rangle, \qquad C_2 = J^2_0 - \frac{1}{2}\{ J_+,
J_- \}, \,\,\, \mu = \frac{1}{4}(\lambda^2 - 1).
\ee
On the other hand, the generators of $hs[\lambda]$ can be written as 
\be
V^s_n = (-1)^{s-1-n} \frac{(n+s-1)!}{(2s-2)!}[ \underbrace{ J_-, \ldots [J_-, [J_- }_{s-1-n\,\,\,\text{terms}},J^{s-1}_+]]], \,\,\, n \geq 2
\ee
while the remaining $V^1_0$ is a central element which we can regard as the identity element. Thus, we can 
write
\be
B[\mu ] = hs[\lambda] \oplus \mathbb{C}
\ee
with $\mathbb{C}$ corresponding to the identity generator.

The algebra's product is called the `lone-star' product. It turns out
to be isomorphic to an ordinary matrix product in an infinite-dimensional matrix representation 
of $hs[\lambda]$ which follows from the following infinite-dimensional matrix 
representation of $sl(2)$.
\bea
(J_0)_{mm} &=& \left( V^2_0 \right)_{mm} = \frac{-\lambda+1}{2} -m, \cr
(J_+)_{m+1,m} &=& \left( V^2_1 \right)_{m+1,m} = -\sqrt{-(\lambda + m)m}, \cr
(J_-)_{mm} &=& \left( V^2_{-1} \right)_{m,m+1} = \sqrt{-(\lambda + m)m}. 
\eea
where $m=1,2,\ldots, \infty$. There is a certain trace operation (see \cite{ProbeCFT,Khesin}) that one can
define for these matrices which reads
\bea
\text{Tr} X &=& \frac{1}{-\lambda} \text{lim}_{M \rightarrow - \lambda} \sum^M_{j=1} X_{jj},\,\,\,
\text{Tr} \{ J_+, J_- \} = -\frac{1}{3} (\lambda^2 - 1 ), \cr
\{J_+, J_- \}_{traceless} &=& \frac{1}{3}\{ J_-, J_+ \} + \frac{4}{3}J^2_0 = \{ J_+, J_- \} + \frac{1}{3} (\lambda^2 - 1) \mathbf{1}.
\eea
Below, we list down some BCH identities which are crucial in helping us navigate through the matrix
algebra. 
\bea
\label{braidzero}
e^{c_+ J_+ + c_- J_-} J_0 e^{-c_+ J_+ - c_- J_-} &=& \textrm{cos}(2\sqrt{c_+ c_-}) J_0 + \frac{\textrm{sin}(2\sqrt{c_+ c_-}) }{2\sqrt{c_+ c_-}} \left( c_+ J_+ - c_- J_-  \right),\\
\label{braidplus}
e^{c_+ J_+ + c_- J_-} J_+ e^{-c_+ J_+ - c_- J_-} &=&  J_+ - \frac{c_-}{\sqrt{c_+ c_-}} \textrm{sin} (2\sqrt{c_+c_-}) J_0  - \frac{1-\textrm{cos}(2\sqrt{c_+ c_-}) }{2c_+ } \left( c_+ J_+ - c_- J_-  \right),\cr 
\,&\,&\,\,\,\\
\label{braidminus}
e^{c_+ J_+ + c_- J_-} J_- e^{-c_+ J_+ - c_- J_-} &=&  J_- + \frac{c_+}{\sqrt{c_+ c_-}} \textrm{sin} (2\sqrt{c_+c_-}) J_0  + \frac{1-\textrm{cos}(2\sqrt{c_+ c_-}) }{2c_- } \left( c_+ J_+ - c_- J_-  \right)\cr
\,&\,&\,\,\,\\
\label{disentanglement}
e^{(\lambda_+ J_+ - \lambda_- J_-)} &=& e^{\Lambda_+ J_+} e^{-\textrm{ln} \Lambda_3 J_0} e^{-\Lambda_- J_-}
\eea
where 
$$
\Lambda_3 = \textrm{sech}^2 \sqrt{\lambda_+ \lambda_-},\qquad
\Lambda_\pm = \frac{\lambda_\pm}{\sqrt{\lambda_+ \lambda_-}}\textrm{tanh}\sqrt{\lambda_+ \lambda_-}.
$$
As we shall see later on, we need to take matrix elements of the following form 
$$
e^{\epsilon J_-} J_{k_1} J_{k_2} \ldots J_{k_n} e^{\beta J_+}. 
$$
Let's begin with an illustrative class of examples and consider  $\left[ e^{\epsilon J_-} J_k J_m e^{\beta J_+} \right]_{11}$. Writing 
$
e^{\epsilon J_-} J_k J_m e^{\beta J_+}  =  
e^{\epsilon J_-} J_k e^{-\epsilon J_-} e^{\epsilon J_-} e^{\beta J_+}
e^{-\beta J_+} J_m e^{\beta J_+}  
\equiv \sum_{i,j} \epsilon_k^i \beta_m^j  J_i e^{\epsilon J_-} e^{\beta J_+} J_j,
$ we then find
\be
\left[ e^{\epsilon J_-} J_k J_m e^{\beta J_+} \right]_{11}= \sum_{i,j} \epsilon_k^i \beta_m^j  \left [J_i e^{\epsilon J_-} e^{\beta J_+} J_j \right]_{11},
\ee
where 
\be
\epsilon_i^k \sim \left( \begin{array}{ccc} 1 & 0 & -\epsilon \\ -2\epsilon & 1 & \epsilon^2 \\ 0 & 0 & 1 \end{array} \right), \qquad \beta_i^k \sim \left( \begin{array}{ccc} 1 & -\beta & 0 \\ 0 & 1 & 0 \\ -2\beta & \beta^2 & 1 \end{array} \right)
\ee
and 
\bea
\left[  J_0 e^{\epsilon J_-} e^{\beta J_+} J_0 \right]_{11} &=& \frac{1}{4}(1-\beta \epsilon)^{-(\lambda + 1)} (\lambda+1)^2, \cr
\left[  J_0 e^{\epsilon J_-} e^{\beta J_+} J_+ \right]_{11} &=& \frac{1}{2}\epsilon(1-\beta \epsilon)^{-(\lambda + 2)} (\lambda+1)^2, \cr
\left[  J_- e^{\epsilon J_0} e^{\beta J_+} J_+ \right]_{11} &=& \frac{1}{2}\beta(1-\beta \epsilon)^{-(\lambda + 2)} (\lambda+1)^2, \cr
\label{twoJmatrix}
\left[  J_- e^{\epsilon J_-} e^{\beta J_+} J_+ \right]_{11} &=& (1-\beta \epsilon)^{-(\lambda + 3)} (\lambda+1)(1+\beta \epsilon (\lambda + 1)). 
\eea
Then the nine matrix elements can be straightforwardly computed to read 
\bea
\left[ e^{\epsilon J_-} J^2_0 e^{\beta J_+} \right]_{11} &=& \frac{1}{4} {(1 - \beta \epsilon)}^{(-3 - \lambda)} (1 + \lambda ) \left(1 + \
\lambda + \beta^2 \epsilon^2 (1 + \lambda) + 
   2 \beta \epsilon (3 + \lambda) \right) \cr
\left[ e^{\epsilon J_-} J_0 J_+ e^{\beta J_+} \right]_{11} &=& -\frac{1}{2} \epsilon {(1 - \beta \epsilon)}^{(-3 - \lambda)} (1 + \lambda ) \left(3 + \
\lambda + \beta \epsilon (1 + \lambda) \right) \cr
\left[ e^{\epsilon J_-} J_0 J_- e^{\beta J_+} \right]_{11} &=& -\frac{1}{2} \beta {(1 - \beta \epsilon)}^{(-3 - \lambda)} (1 + \lambda ) \left(1 + \
\lambda + \beta \epsilon (3 + \lambda) \right) \cr
\left[ e^{\epsilon J_-} J_+ J_0 e^{\beta J_+} \right]_{11} &=& -\frac{1}{2} \epsilon {(1 - \beta \epsilon)}^{(-3 - \lambda)} (1 + \lambda ) \left(1 + \
\lambda + \beta \epsilon (3 + \lambda) \right) \cr
\left[ e^{\epsilon J_-} J^2_+  e^{\beta J_+} \right]_{11} &=& \epsilon^2 {(1 - \beta \epsilon)}^{(-3 - \lambda)} (1 + \lambda )(2 + \lambda) \cr
\left[ e^{\epsilon J_-} J_+ J_- e^{\beta J_+} \right]_{11} &=& \beta \epsilon {(1 - \beta \epsilon)}^{(-3 - \lambda)} (1 + \lambda ) (1+\beta \epsilon + \lambda)\cr
\left[ e^{\epsilon J_-} J_- J_0 e^{\beta J_+} \right]_{11} &=& -\frac{1}{2} \epsilon {(1 - \beta \epsilon)}^{(-3 - \lambda)} (1 + \lambda ) \left(3 + \
\lambda + \beta \epsilon (1 + \lambda) \right) \cr
\left[ e^{\epsilon J_-} J_- J_+ e^{\beta J_+} \right]_{11} &=& {(1 - \beta \epsilon)}^{(-3 - \lambda)} (1 + \lambda ) \left(1 + \beta \epsilon (1 + \lambda) \right) \cr
\left[ e^{\epsilon J_-} J_-^2 e^{\beta J_+} \right]_{11} &=&  \beta^2 {(1 - \beta \epsilon)}^{(-3 - \lambda)} (1 + \lambda )(2 + \lambda).
\eea
For other matrix elements containing higher powers of the $sl(2)$ generators, the problem reduces to finding
$\left[ J_{k_1} J_{k_2} \ldots J_{k_m} e^{\epsilon J_-} e^{\beta J_+} J_{l_1} J_{l_2} \ldots J_{l_n} \right]_{11}$, which can be derived recursively from \eqref{twoJmatrix} by invoking the $sl(2)$ algebra and taking suitable derivatives. 
For example, we find the following formulae useful. 
\be
\label{traceformula1}
\left[ e^{\Lambda_- J_-}e^{-\Lambda_+ J_+}J_+^n \right]_{11} = \frac{\Gamma (\lambda+n+1) }{\Gamma (\lambda+1)} \Lambda^n_- \left( 1 + \Lambda_- \Lambda_+  \right)^{-(\lambda+n+1)},
\ee
\be
\label{traceformula2}
\left[ J_-^n e^{\Lambda_- J_-}e^{-\Lambda_+ J_+} \right]_{11} = \frac{\Gamma (\lambda+n+1) }{\Gamma (\lambda+1)} \left( -\Lambda_+ \right)^n   \left( 1 + \Lambda_- \Lambda_+  \right)^{-(\lambda+n+1)},
\ee
which can be derived by taking derivatives of 
\be
\label{traceformula1}
\left[ e^{\Lambda_- J_-}e^{-\Lambda_+ J_+} \right]_{11} = \frac{\Gamma (\lambda+1) }{\Gamma (\lambda+1)}  \left( 1 + \Lambda_- \Lambda_+  \right)^{-(\lambda+1)}.
\ee
Back to the bulk-boundary propagator, in such an infinite-dimensional matrix representation, 
the highest weight state $c$ is a projector that reads $c=\text{Diag} \left(1,0,0,\ldots \right)$. 
Substituting this into \eqref{linearizedscalar}, it reduces to the matrix element
\be
\label{scalarfield1}
\Phi (\rho, z,\bar{z}) = e^{\Delta \rho} \langle 1 \vert e^{\bar{\Lambda}} e^{-\Lambda} \vert 1\rangle
\ee
We can then attempt to use the results of this section to evaluate this explicitly. 
Below, we will study the case where the Chern-Simons connection can be written as a perturbation
series in the higher-spin chemical potential. This pertains to classical bulk solutions which can be viewed as
higher-spin deformations of solutions in pure gravity. In particular, in this paper, we wish to compute the scalar propagators in the background of higher-spin deformations of the BTZ black hole of which Chern-Simons connection can be computed order by order following a prescription in \cite{KrausPartition} to ensure consistent gravitational thermodynamics for them. The crucial aspect of computational feasibility lies in the fact that the Chern-Simons connection for the background contains only terms linear in the $sl(2)$ generators. The higher-spin corrections can then be expressed (at least in terms of integrals) to any order in perturbation theory.

\subsection{Some comments on the perturbation series for higher-spin deformations}

Let us consider perturbation about an ordinary gravitational background for which we denote its Chern-Simons connection as $\Lambda^{(0)}$. (Henceforth, all bracketed superscripts refer to orders in perturbation theory.) We denote scalar bulk-boundary propagator as $\Phi^{(0)}$, and the higher-spin chemical potential as $\alpha$. A general higher-spin deformation involves another non-vanishing chemical potential $\bar{\alpha}$ that couples independently to the anti-holomorphic Chern-Simons connection. For simplicitly we will set it to be zero. The results that we obtain in this paper can be generalized to a non-zero $\bar{\alpha}$ straightforwardly. Expanding about $\alpha$, we write
$
\Phi (\rho, z,\bar{z}) = \Phi^{(0)} (\rho, z,\bar{z})+ \sum_{n=1}^{\infty} \Phi^{(n)} (\rho, z,\bar{z}),\,\,
\Lambda (\rho, z, \bar{z}) = \Lambda^{(0)} (\rho, z, \bar{z}) + \sum_{n=1}^{\infty} \alpha^n 
\Lambda^{(n)} (\rho, z, \bar{z}), \,\,\bar{\Lambda} (\rho, z, \bar{z})= \bar{\Lambda}^{(0)} (\rho, z, \bar{z}).
$
Up to second order, the propagator reads 
\bea
\Phi (\rho, z,\bar{z}) &=& e^{(1+\lambda)\rho} \left[  e^{\bar{\Lambda}} e^{-\Lambda} \right]_{11} \cr 
&=& e^{(1+\lambda)\rho} \left[  e^{\bar{\Lambda}} e^{- \left(
\Lambda^0 + \alpha \Lambda^1 + \alpha^2 \Lambda^2 \right)
} \right]_{11} \cr
&=& e^{(1+\lambda )\rho} \left[ e^{\bar{\Lambda}} \left(  e^{-\Lambda^0} 
+\alpha \frac{d}{d\alpha} e^{-\Lambda}\vert_{\alpha=0} + \alpha^2 \frac{d^2}{d\alpha^2} e^{-\Lambda}\vert_{\alpha = 0} \right) \right]_{11} +\mathcal{O}(\alpha^3)
\eea
where the derivatives of the matrix exponentials read
\be
\frac{d}{d\alpha} e^{-\Lambda} = -\int^1_0 dw\, e^{-w \Lambda} \Lambda
e^{w\Lambda} e^{-\Lambda}
\Rightarrow
\frac{d}{d\alpha} e^{-\Lambda}\vert_{\alpha = 0} = -\int^1_0 dw\, e^{-w \Lambda^0} \Lambda^1
e^{w\Lambda^0} e^{-\Lambda^0}.
\ee
\bea
\frac{d^2}{d\alpha^2} e^{-\Lambda}\vert_{\alpha=0} &=& \int^1_0 ds\,\, \Bigg( -2e^{-s\Lambda^0}\,\Lambda^2
e^{s\Lambda^0} e^{-\Lambda^0} - \left(  \frac{d}{d\alpha}e^{-s\Lambda}\vert_{\alpha = 0} \right) \Lambda^1 e^{s\Lambda^0}e^{-\Lambda^0} \cr
&&- e^{-s\Lambda^0} \Lambda^1 \frac{d}{d\alpha} e^{(s-1)\Lambda} \vert_{\alpha = 0} \Bigg).
\eea

In this paper, we will be primarily interested in the higher-spin black holes first discussed in \cite{KrausPartition}. Now, there are other higher-spin black holes alluded to in \cite{OtherBH1} and \cite{OtherBH2}. These were shown to be gauge-equivalent to the ones of interest in this work. 
Recall that the Chern-Simons gauge fields are pure gauges $A=g^{-1} dg, \bar{A} = \bar{g}^{-1} d\bar{g}$.
With a further gauge transformation via $(h,\bar{h})$, the physical scalar field reads
\be
\Phi (\rho, z, \bar{z} ) = e^{\Delta \rho} \left[ e^{\bar{\Lambda}_\rho} \bar{h} h^{-1} e^{-\Lambda_\rho} \right]_{11}. 
\ee
In \cite{OtherBH2}, $(h,\bar{h})$ were derived up to first-order in the higher-spin chemical potentials. One can similarly compute the scalar propagator in the higher-spin black holes of \cite{OtherBH1} and \cite{OtherBH2} perturbatively by using the techniques in this section, since at each order, the expressions for $(h,\bar{h})$ are just some polynomials in $J$'s.

\subsection{Simple examples: Bulk-boundary propagator of the BTZ black hole and chiral deformations of
$AdS_3$}

Before we compute the propagator in the background of a higher-spin black hole, it is
instructive to use the above procedure to compute the propagator in the background of
the BTZ black hole. The Chern-Simons connections in this case read
\be
a=\left( J_+ + \frac{1}{4\tau^2}J_- \right) dz, \bar{a}=\left( J_- + \frac{1}{4\bar{\tau}^2}J_+ \right) d\bar{z}. 
\ee
This gives rise to the BTZ metric in Euclidean signature
\be
ds^2 = d\rho^2 + \frac{2\pi}{k}\left( \mathcal{L} dz^2 + \bar{\mathcal{L}}  d\bar{z}^2 \right) + 
\left( e^{2\rho} + \left( \frac{2\pi}{k} \right)^2 \mathcal{L} \bar{\mathcal{L}}e^{-2\rho} \right) dz d\bar{z},
\ee
where $\mathcal{L}=-k/(8\pi \tau^2), \bar{\mathcal{L}} = -k/(8\pi \bar{\tau}^2)$ are the left and right moving components of the boundary stress-momentum tensor with $\tau$ being the modular parameter of the Euclidean boundary torus. The black hole horizon 
is located at $\rho_{hori.} = -\textrm{log}(4\tau \bar{\tau})/2$. The scalar bulk-boundary propagator
is known to be 
\be
\label{btzprop}
G (\rho, z, \bar{z}) = \left(  \frac{e^{-\rho}}{e^{-2\rho}\cos(\frac{z}{2\tau}) \cos(\frac{\bar{z}}{2\bar{\tau}}) + 4\tau \bar{\tau} \sin(\frac{z}{2\tau}) \sin(\frac{\bar{z}}{2\bar{\tau}})    }  \right)^{1+ \lambda}.
\ee
Let's check that \eqref{btzprop} can be expressed as a matrix element. This will serve as a good consistency check of 
our computational approach. 
For the BTZ, the connections imply that we need to compute the matrix element
\be
\label{phibtz}
\Phi^{BTZ} (\rho, z,\bar{z}) = e^{\rho (1 + \lambda)} \left[ e^{a_+ J_+ - a_- J_-}
e^{\tilde{a}_+ J_+ - \tilde{a}_- J_-}   \right]_{11} 
\ee
where 
\be
a_- = -e^\rho \bar{z}, \,\,\,a_+ = \frac{e^{-\rho}\bar{z}}{4\bar{\tau}^2},\,\,\,\tilde{a}_+
= -e^\rho z, \,\,\,\tilde{a}_- = \frac{e^{-\rho}z}{4\tau^2}.
\ee
From \eqref{disentanglement}, and the relation $(J_0)_{11} = -\frac{1+\lambda}{2}$,
we find \eqref{phibtz} to simplify to \eqref{btzprop}. 

We should mention another worked example for which the propagator can be computed to all orders in the
higher-spin potential and which was already done in \cite{ProbeCFT} - the chiral spin-3 perturbation of the $AdS_3$ vacuum. The connections read
\be
\label{spin3perturbation}
a=J_+ dz - \mu V^3_2 d\bar{z},\qquad \bar{a} = J_- d\bar{z}
\ee
where $\mu$ is constant. From the viewpoint of holography, this is the dual to a deformation of the boundary
by a dimension (3,0) operator $\mathcal{W}(z)$ with constant coupling, schematically, $\delta S_{CFT} = \mu
\int d^2 z \mathcal{W}(z)$. The higher-spin black hole solution with chiral charge is a finite temperature version of
this solution. For this background, we note that
\be
\Lambda = e^{\rho} z J_+ - \mu e^{2\rho} \bar{z} J^2_+, \qquad \bar{\Lambda} = e^\rho \bar{z},
\ee
and that the propagator can then be computed straightforwardly as
\bea
\label{scalarprop}
\Phi (\rho, z,\bar{z}) &=& e^{(1+\lambda)\rho} \left[ e^{e^\rho \bar{z} J_-} e^{-e^\rho z J_+ + \mu e^{2\rho} \bar{z} J^2_+} \right]_{11} \cr
&=& e^{(1+\lambda)\rho} \left[ \sum_{s=0}^{\infty} e^{e^\rho \bar{z} J_-}e^{-e^\rho z J_+} \frac{(\mu e^{2\rho}\bar{z}J^2_+)^s}{s!}   \right]_{11} \cr
&=&e^{(1+\lambda)\rho}\sum_0^{\infty} \frac{(\mu \bar{z}^3 e^{4\rho})^s}{s!} \frac{(\lambda + 2s)!}{\lambda! (1+e^{2\rho}\bar{z}z)^{1+\lambda+2s}  } \cr
&=& \left(  \frac{e^{\rho}}{1+e^{2\rho} |z|^2} \right)^{1+ \lambda} \sum_{n=0}^{\infty} c_n
\left( \frac{\mu \bar{z}^3 e^{4\rho}}{(1+|z|^2e^{2\rho})^2} \right)^n, \,\,\,
c_n \equiv \prod_{i=1}^n (i + \lambda),
\eea
as first derived in \cite{ProbeCFT}. 
However, for the spin-3 black hole with non-zero temperature, the propagator at arbitrary $\lambda$ wasn't computed in a similar fashion. We now proceed to study this case, demonstrating that the techniques in this Section are sufficient for us to compute the bulk-boundary propagator.

\section{A derivation of the scalar propagator in a higher-spin black hole background}
\label{sec:Prop}
In this section, we will derive the bulk-boundary propagator and by taking $\rho \rightarrow \infty$, the boundary two-point functions in the background of spin-3 and spin-4 black holes. We will give explicit 
expressions up to second order for the spin-3 black hole and first order for the spin-4 case.  

\subsection{Spin-3 black hole}
For the spin-3 black hole, the $hs[\lambda] \times hs[\lambda]$ Chern-Simons connection reads 
\bea
a_z &=& V^2_1 - \frac{2\pi \mathcal{L}}{k}V^2_{-1} - \frac{\pi \mathcal{W}}{2k}V^3_{-2} + \mathcal{U} V^4_{-3} + \ldots \\
a_{\bar{z}} &=& -\frac{\alpha}{\bar{\tau}} \left( a_z \ast a_z \right)\vert_{traceless},
\eea
with the charges being (up to second order in $\alpha$)
\bea
\mathcal{L} &=& -\frac{k}{8\pi \tau^2} + \frac{k}{24\pi \tau^6} (\lambda^2 - 4) \alpha^2 + \mathcal{O}(\alpha^4), \\
\mathcal{W} &=& - \frac{k}{3\pi \tau^5} \alpha + \mathcal{O}(\alpha^3),\\
\mathcal{U} &=& \frac{7}{36\tau^8} \alpha^2 + \mathcal{O}(\alpha^4).
\eea
We refer the reader to \cite{KrausPartition} for the derivation. Up to second order,
\bea
\Lambda^0 &=& z( e^\rho J_+ + e^{-\rho} \frac{1}{4\tau^2} J_- = ZT J_+ + \frac{Z}{4T}J_-, \\
\Lambda^1 &=& z \left( \frac{e^{-2\rho}}{6\tau^5} J^2_- \right) - 
\frac{\bar{z}}{\bar{\tau}} \left( e^{2\rho} J^2_+ + \frac{e^{-2\rho}}{16\tau^4}J^2_- + \frac{1}{4\tau^2}\{ J_+, J_- \}_{traceless} \right)\cr
&=& e^{2\rho} \left( Z\left( \frac{1}{6T^4} J^2_- \right) - \bar{Z} \left( J^2_+ + \frac{1}{16T^4}J^2_- + \frac{1}{4T^2}\left(  \frac{1}{3}\{J_+, J_-\} + \frac{4}{3}J^2_0 \right) \right) \right), \\
\Lambda^2 &=& z\left(  \frac{7}{36 \tau^8} e^{-3\rho} J^3_- - \frac{(\lambda^2 - 4)}{12\tau^6} e^{-\rho}J_- \right) - \frac{\bar{z}}{\bar{\tau}} \left( \frac{e^{-\rho}}{6\tau^5} \{ J_+, J^2_- \} + \frac{e^{-3\rho}}{12\tau^7} J^3_- \right) \cr
&=& e^{4\rho} \left[  -\frac{(\lambda^2 - 4)Z}{12T^5} J_- + \left( \frac{7Z-3\bar{Z}}{36T^7}   \right) J^3_- - \frac{\bar{Z}}{6T^5} \{J_+, J^2_- \} \right],\cr
\bar{\Lambda} &=&  \bar{z} \left( e^\rho J_- + e^{-\rho} \frac{1}{4\bar{\tau}^2} J_+ \right) = \bar{Z}\bar{T} J_- + \frac{\bar{Z}}{4\bar{T}}J_+, 
\eea
where we have defined 
$$
T \equiv \tau e^\rho, \qquad \bar{T} = \bar{\tau} e^\rho,\qquad Z= \frac{z}{\tau}, \qquad \bar{Z} = \frac{\bar{z}}{\bar{\tau}},
$$
and have set the anti-holomorphic chemical potential to vanish.
Using \eqref{braidzero}, \eqref{traceformula1} and \eqref{traceformula2} in Section \ref{sec:Prelim}, we can write 
\be
\label{braiding}
e^{-s\Lambda^0} J_i e^{s\Lambda^0} = \sum_i \mathcal{C}^k_i (s,Z) J_k,
\ee
where, representing the coefficients $\mathcal{C}^k_i (s,Z)$ as matrix elements of $\mathcal{C}$, $i=\{1,2,3\}=\{0,+,-\} $
\be
\mathcal{C} = \left( \begin{array}{ccc} \cos (sZ) & -T \sin (sZ) & \frac{\sin (sZ)}{4T} \\
\frac{\sin (sZ)}{2T} & \frac{1+\cos (sZ)}{2} & -\frac{\cos (sZ) -1}{8T^2} \\
-2T \sin (sZ) & 2T^2 ( 1 -\cos (sZ) ) & \frac{(1+\cos (sZ))}{2}
\end{array} \right).
\ee
To complete the calculation at each order, using \eqref{disentanglement}, we are then left with linear combinations of the matrix elements
\be
\left( \sec \frac{\bar{Z}}{2} \right)^{1+\lambda}\left( \sec \frac{Z}{2} \right)^{1+\lambda} \times
\left[ e^{\epsilon J_-} J_i J_j J_k \ldots e^{\beta J_+} \right]_{11}
\ee
where $\epsilon = 2\bar{T} \tan \frac{\bar{Z}}{2}, \beta = -2T \tan \frac{Z}{2}$.

\subsubsection{First order correction}
For first-order correction, we need to compute
\bea
-\int^1_0 ds\,\,\, e^{-s\Lambda^0} \Lambda^1\, e^{s\Lambda^0} &=&-e^{2\rho} \sum_{i,j} \mathcal{D}^{ij}(Z,\bar{Z}) \int^1_0 ds\,\,\, \mathcal{C}^k_i (s,Z) \mathcal{C}^m_j (s,Z) J_k J_m \cr
&\equiv& -\int^1_0 ds\,\mathcal{B}^{km}(s,Z,\bar{Z}) J_k J_m \equiv -\mathcal{B}^{km}(Z,\bar{Z}) J_k J_m, 
\eea
where 
\be
\mathcal{D}^{33}=\frac{Z}{6T^4}-\frac{\bar{Z}}{16T^4}, \mathcal{D}^{22} = -\bar{Z}, 
\mathcal{D}^{23} = \mathcal{D}^{32} = -\frac{\bar{Z}}{12T^2}, \mathcal{D}^{11} =  -\frac{\bar{Z}}{3T^2}.
\ee
Then, we simply take the matrix elements
\be
\left[   e^{\bar{\Lambda}^0}  \mathcal{B}^{km}(Z,\bar{Z}) J_k J_m  e^{-\Lambda^0} \right]_{11}
=\left( \sec \frac{\bar{Z}}{2} \right)^{1+\lambda}\left( \sec \frac{Z}{2} \right)^{1+\lambda} \times
\mathcal{B}^{km}(Z,\bar{Z})  \left[ e^{\epsilon J_-} J_k J_m e^{\beta J_+} \right]_{11}.
\ee
Let us now determine the coefficients $\mathcal{B}^{km}(s,Z,\bar{Z}), \mathcal{B}^{km}(Z,\bar{Z}),$. They read 
\be
\mathcal{B}^{km}(s,Z,\bar{Z}) = \sum_{i,j} \mathcal{D}^{ij}(Z,\bar{Z}) \, \mathcal{C}^k_i (s,Z) \mathcal{C}^m_j (s,Z) 
\ee
and $\mathcal{B}\left( Z, \bar{Z} \right) \equiv \int^1_0\,ds\, \mathcal{B}\left(s, Z, \bar{Z} \right)$ reads in matrix form
\be
\left( \begin{array}{ccc} 
\frac{ Z - \bar{Z} - \cos (Z) \sin (Z)   }{3T^2}  & -\frac{4}{3T}\sin^4 (\frac{Z}{2})
& -\frac{(\cos (Z) +3)}{6T^3} \sin^2 (\frac{Z}{2}) \\
-\frac{4}{3T}\sin^4 (\frac{Z}{2}) & \frac{1}{6} (6 (Z - \bar{Z}) - 8 \sin(Z) + \sin(2 Z))
& \frac{ Z - \bar{Z} - \cos (Z) \sin (Z)   }{12T^2} \\
-\frac{(\cos (Z) +3)}{6T^3} \sin^2 (\frac{Z}{2}) & \frac{ Z - \bar{Z} - \cos (Z) \sin (Z)   }{12T^2} & \frac{1}{48 T^4} \left( 3(Z-\bar{Z}) + (4 + \cos (Z)) \sin (Z) \right) \end{array} \right).
\ee
We now have all the ingredients to compute the first-order correction. Assembling everything, we have 
\bea
\Phi^{(1)}(\rho,z,\bar{z}) &=& \Phi^{BTZ}(\rho,z,\bar{z}) \frac{(\lambda+1)(\lambda +2)}{ 12\tau^2\left(
\cos (\frac{Z}{2})\cos (\frac{\bar{Z}}{2}) + 4T \bar{T} \sin (\frac{Z}{2})\sin (\frac{\bar{Z}}{2}) 
\right)^{2}} \cr
&\times& \Bigg[ \cos^2 \frac{\bar{Z}}{2} \left( \sin Z + (2-\cos Z)(\bar{Z} - Z) \right) \cr
&&-4T\bar{T} \sin \bar{Z}  \left( 2(1- \cos Z) + \sin Z (\bar{Z} - Z) \right) \cr
&&+16 T^2 \bar{T}^2 \sin^2 \frac{\bar{Z}}{2} \left( 3 \sin Z + (2+\cos Z)(\bar{Z} - Z) \right) \Bigg].
\eea
This was derived in \cite{ProbeCFT} in a rather different spirit\footnote{We thank Eric Perlmutter for a couple of helpful email exchanges on this issue when \cite{ProbeCFT} first appeared.}, 
where a generalized Klein-Gordan 
equation was extracted by decoupling various components of the linearized master field equation \eqref{MasterKG}. It was not clear how to systematically seek solutions to the PDE that emerge but the conjectured dual CFT correlator that should be
identified with the infinite $\rho$ limit was calculated, and presumably this allowed them to arrive at a plausible ansatz that could be checked to solve the PDE. We should mention that another seemingly different method of 
obtaining the first-order bulk-boundary propagator was presented in 
\cite{OtherBH2} where a relation to the spectrum of quasi-normal modes was emphasized. 

The boundary two-point function can be obtained by taking the infinite $\rho$ limit, which yields 
\be
\lim_{\rho \rightarrow \infty} \Phi^{(1)} (\rho,z,\bar{z}) =\lim_{\rho \rightarrow \infty} \Phi^{BTZ} (\rho, z,\bar{z})\times \frac{(1+\lambda)(2+\lambda)}{6 \tau^2} 
\frac{2 \sin Z + (2 + \cos Z)(\bar{Z} - Z)}{2\sin^2 \frac{Z}{2}}, 
\ee
where $\Phi^{BTZ}$ denotes the corresponding correlator for the BTZ black hole. This was first derived in \cite{ProbeCFT} where it was nicely related to the CFT three-point function $\langle W(z_1, \bar{z}_1) \bar{\phi} (z_2, \bar{z}_2) \phi (z_3,\bar{z}_3) \rangle$. We will not review the relevant CFT derivation here, but in Section \ref{sec:Boundary} and in
Appendix \ref{sec:AppB}
we will perform a CFT derivation of the boundary two-point function in first and second-orders of the spin-4 chemical potentials respectively.

\subsubsection{The second-order correction}
We now proceed to computing the second-order correction which involves calculating
\bea
\frac{d^2}{d\alpha^2} e^{-\Lambda}\vert_{\alpha=0} &=& -\int^1_0 ds\,\, \Bigg( 2e^{-s\Lambda^0}\,\Lambda^2
e^{s\Lambda^0} e^{-\Lambda^0} + \left(  \frac{d}{d\alpha}e^{-s\Lambda}\vert_{\alpha = 0} \right) \Lambda^1 e^{s\Lambda^0}e^{-\Lambda^0} \cr
\label{3secondorder}
&&+ e^{-s\Lambda^0} \Lambda^1 \frac{d}{d\alpha} e^{-(1-s)\Lambda} \vert_{\alpha = 0} \Bigg). 
\eea
We note that 
\be
\frac{d}{d\alpha} e^{-s\Lambda}\vert_{\alpha = 0} 
=\mathcal{B}^{km} \left( -sZ, -s\bar{Z} \right) J_k J_m e^{-s\Lambda^0}
\ee
\be
\frac{d}{d\alpha} e^{(s-1)\Lambda}\vert_{\alpha = 0} 
=\mathcal{B}^{km} \left( (s-1)Z, (s-1)\bar{Z}  \right) J_k J_m e^{(s-1)\Lambda^0}.
\ee
Thus, the second and third terms in \eqref{3secondorder} can be expressed respectively as
\be
\left( -\int^1_0 ds\,\, \mathcal{B}^{km}(-sZ, -s\bar{Z} ) B^{ij}(s, Z, \bar{Z} ) \right) J_k J_m J_i J_j e^{-\Lambda^0},
\ee
\be
\left( -\int^1_0 ds\,\, \mathcal{C}_r^i (s,Z) \mathcal{C}_s^j (s,Z) \mathcal{B}^{rs}\left( (s-1)Z, (s-1)\bar{Z} \right) \mathcal{B}^{km} (s,Z,\bar{Z}) \right)  J_k J_m J_i J_j e^{-\Lambda^0}.
\ee
We are then left with computing matrix elements of the form 
\be
\left[ e^{\epsilon J_-} J_k J_m J_i J_j e^{\beta J_+} \right]_{11}
= \epsilon_k^p \epsilon_m^q \beta_i^r \beta_j^s \,\, \left[ J_p J_q e^{\epsilon J_-}
e^{\beta J_+}J_r J_s \right]_{11},
\ee
which can be handled using the techniques presented in Section \ref{sec:Prelim}. 
On the other hand, the first term of \eqref{3secondorder} takes into account the second-order correction to 
the Chern-Simons connection
\be
\Lambda^2 =  e^{4\rho} \left[  -\frac{(\lambda^2 - 4)Z}{12T^5} J_- + \left( \frac{7Z-3\bar{Z}}{36T^7}   \right) J^3_- - \frac{\bar{Z}}{6T^5} \{J_+, J^2_- \} \right],
\ee
and upon using similar techniques, we can simplify $\int^1_0 ds\,\,e^{-s\Lambda^0} \Lambda^2 e^{s\Lambda^0}$ to read
\bea
\int^1_0 ds\, && \left( -\frac{(\lambda^2 - 4)Z}{12T^5} \right) \mathcal{C}^i_- (s,Z) J_i + 
\left( \frac{7Z-3\bar{Z}}{36T^7}   \right)\sum_{i,j,k} \mathcal{C}^i_- (s) \mathcal{C}^j_- (s,Z) \mathcal{C}^k_- (s,Z) J_i J_j J_k \cr
&&- \left( \frac{\bar{Z}}{6T^5} \right) \left[ 2\sum_{i,j} \mathcal{C}^i_0 (s,Z) \mathcal{C}^j_- (s,Z)
J_i J_j + \sum_{i,j,k} \mathcal{C}^i_- (s,Z) \mathcal{C}^j_{\{+} (s,Z) \mathcal{C}^k_{-\}} (s,Z)
J_i J_j J_k \right]. \nonumber 
\eea
After performing the $s$-integration, we are again left with matrix elements of the form
$$
\left[ e^{\epsilon J_-} J_k J_m J_i  e^{\beta J_+} \right]_{11}
= \epsilon_k^p \epsilon_m^q \beta_i^r  \,\, \left[ J_p J_q e^{\epsilon J_-}
e^{\beta J_+}J_r  \right]_{11},\qquad 
\left[ e^{\epsilon J_-} J_k J_m  e^{\beta J_+} \right]_{11}
= \epsilon_k^p  \beta_m^r  \,\, \left[ J_p e^{\epsilon J_-}
e^{\beta J_+}  J_r \right]_{11}.
$$
Assembling all the terms together, after some simplification, we found that the second-order contribution to the bulk-boundary propagator can be expressed as
\be
\Phi^{(2)}(\rho,z,\bar{z})= \Phi^{BTZ}(\rho,z,\bar{z}) \frac{e^{4\rho} (\lambda +1) (\lambda+2) \sum_{i=m}^4 f^{(m)}(Z,\bar{Z}) (T\bar{T})^m }{T^4 \left( \cos \left( \frac{Z}{2} \right) \cos \left( \frac{\bar{Z}}{2} \right) + 4T\bar{T} \sin \left( \frac{Z}{2} \right) \sin \left( \frac{\bar{Z}}{2} \right)   \right)^4}
\ee
where the $f's$ read
\bea
f^{(0)}&=&-\frac{1}{144} \left(\lambda + 4\right)\left(\lambda -13\right) \cos\left[\frac{\bar{Z}}{2}\right]^4 \sin[Z]^2 \cr
&-&\frac{1}{72} \left((-4+\lambda ) \cos\left[\frac{\bar{Z}}{2}\right]^4 (-2 (4+\lambda )+(1+\lambda ) \cos[Z]) \sin[Z]\right) (Z-\bar{Z}) \cr
&-&\frac{1}{288} \left(\cos\left[\frac{\bar{Z}}{2}\right]^4 \left(9 \left(14+7 \lambda +\lambda ^2\right)-4 \left(\lambda + 2\right)\left(2\lambda + 11\right) \cos[Z]+\left( \lambda +2 \right)\left( \lambda + 1 \right) \cos[2 Z]\right)\right) (Z-\bar{Z})^2, \nonumber \\
f^{(1)}&=&-\frac{1}{9} \Bigg(
\cos\left[\frac{\bar{Z}}{2}\right]^3 \left(-2 \left(10+5 \lambda +\lambda ^2\right) \cos\left[\frac{Z}{2}\right]+2 \left(10+5 \lambda +\lambda ^2\right) \cos\left[\frac{3 Z}{2}\right]\right) \sin\left[\frac{\bar{Z}}{2}\right] \sin\left[\frac{Z}{2}\right] \cr
&-&2 \left(\cos\left[\frac{\bar{Z}}{2}\right]^3 \Bigg(-94-39 \lambda -5 \lambda ^2
+\left(14+15 \lambda +\lambda ^2\right) \cos[Z]\Bigg) \sin\left[\frac{\bar{Z}}{2}\right] \sin\left[\frac{Z}{2}\right]^2\right) (Z-\bar{Z}) \cr
&+&\cos\left[\frac{\bar{Z}}{2}\right]^3 \left(-3 \left(14+9 \lambda +\lambda ^2\right) \cos\left[\frac{Z}{2}\right]+\left(2+3 \lambda +\lambda ^2\right) \cos\left[\frac{3 Z}{2}\right]\right) \sin\left[\frac{\bar{Z}}{2}\right] \sin\left[\frac{Z}{2}\right] (Z-\bar{Z})^2 \Bigg), \nonumber \\
f^{(2)} &=& \frac{1}{9} \left(-196-99 \lambda -11 \lambda ^2+\left(76+45 \lambda +5 \lambda ^2\right) \cos[Z]\right) \sin^2\bar{Z} \sin\left[\frac{Z}{2}\right]^2 \cr
&-&\frac{1}{9} \left(\left(-4 \left(22+15 \lambda +2 \lambda ^2\right)+\left(28+33 \lambda +5 \lambda ^2\right) \cos[Z]\right) \sin[\bar{Z}]^2 \sin[Z]\right) (Z-\bar{Z}) \cr
&+&\frac{1}{12} \left(-3 \left(14+7 \lambda +\lambda ^2\right)+\left(2+3 \lambda +\lambda ^2\right) \cos[2 Z]\right) \sin[\bar{Z}]^2 (Z-\bar{Z})^2, \\
f^{(3)} &=& \frac{32}{3} \left(8+6 \lambda +\lambda ^2\right) \sin\left[\frac{\bar{Z}}{2}\right]^2 \sin[\bar{Z}] \sin\left[\frac{Z}{2}\right]^2 \sin[Z] \cr
&-&\frac{32}{9} \left(\cos\left[\frac{\bar{Z}}{2}\right] \left(74+45 \lambda +7 \lambda ^2+\left(22+27 \lambda +5 \lambda ^2\right) \cos[Z]\right) \sin\left[\frac{\bar{Z}}{2}\right]^3 \sin\left[\frac{Z}{2}\right]^2\right) (Z-\bar{Z}) \cr
&+&\frac{16}{9} (2+\lambda ) \cos\left[\frac{\bar{Z}}{2}\right] (11+2 \lambda +(1+\lambda ) \cos[Z]) \sin\left[\frac{\bar{Z}}{2}\right]^3 \sin[Z] (Z-\bar{Z})^2, \\
f^{(4)} &=&-\frac{32}{3} \left((4+\lambda ) (13+3 \lambda +(5+3 \lambda ) \cos[Z]) \sin\left[\frac{\bar{Z}}{2}\right]^4 \sin\left[\frac{Z}{2}\right]^2\right) \cr
&+&\frac{32}{3} (4+\lambda ) (8+2 \lambda +\cos[Z]+\lambda  \cos[Z]) \sin\left[\frac{\bar{Z}}{2}\right]^4 \sin[Z] (Z-\bar{Z}) \cr
&-&\frac{8}{9} \left(\left(9 \left(14+7 \lambda +\lambda ^2\right)+\left(88+60 \lambda +8 \lambda ^2\right) \cos[Z]+\left(2+3 \lambda +\lambda ^2\right) \cos[2 Z]\right) \sin\left[\frac{\bar{Z}}{2}\right]^4\right) (Z-\bar{Z})^2. \nonumber \\
\eea
Both first- and second-order contributions to the bulk-boundary propagator are evidently periodic under 
$(Z,\bar{Z}) \sim (Z + 2\pi \mathbb{Z}, \bar{Z} + 2\pi \mathbb{Z} )$ which is a simple consistency check
for our results. As first noted in \cite{ProbeCFT}, this comes from the fact that the higher-spin
black hole is defined with reference to that of the BTZ which is a central element of $hs[\lambda]$. From \eqref{eigenvalueConstraint}, we can see that the scalar propagator is periodic as such. Since $T\equiv
\tau e^\rho$, we see that only the term $f^{(4)}(Z,\bar{Z})$ remains relevant for the boundary two-point function which reads 
\be
\lim_{\rho \rightarrow \infty} \Phi^{(2)}(\rho,z,\bar{z}) = \lim_{\rho \rightarrow \infty} \Phi^{BTZ}(\rho, z,\bar{z}) \times
\frac{(\lambda +1)(\lambda + 2)f^{(4)}(Z,\bar{Z}) }{256 \tau^4 \sin^4 \frac{Z}{2} \sin^4 \frac{\bar{Z}}{2}}.
\ee
This expression was derived from the dual scalar correlator in the dual CFT in \cite{ProbeCFT} but the boundary two-point function from the bulk's viewpoint was only computed for some integer values of $\lambda$ and not for arbitrary values as we have derived here. This was extended to arbitrary $\lambda$ in \cite{resum} where the computation relied on the analytic continuation from an integer as well. In both \cite{ProbeCFT} and \cite{resum}, the second-order contribution to the bulk-boundary propagator as computed in this Section was not derived. 

Higher-order contributions to the bulk-boundary propagator are in principle calculable, and the computational feasibility depends on the number of terms of the form $\left[ e^{\epsilon J_-} J_{k_1} J_{k_2} \ldots J_{k_m} e^{\beta J_+} \right]_{11}$ that fall out of the perturbative expansion. As we have seen in the first two orders, the coefficients of these matrix elements involve elementary integration over polynomials of trigonometric functions. Below, we study another higher-spin black hole - that which is deformed by a spin-4 chemical potential. Unfortunately, we are only able to study it explicitly at first order. The second-order term can be expressed easily in closed form just as we have done in the spin-3 case. The problem that arose was the proliferation of terms for which we could not find an obvious way to simplify.

\subsection{Spin-4 black hole}

The higher-spin black hole in $hs[\lambda]$ theory with spin-4 charge was first introduced explicitly in 
\cite{Spin4}. We will however adopt a different normalization for the higher-spin fields. The Chern-Simons connections read
\bea
a_+ &=& J_+ + \left(  \frac{1}{4\tau^2} + \frac{3(\lambda^2 - 9)(\lambda^2 -4) \alpha^2}{400 \tau^8} \right) J_- +
\left(  \frac{\alpha}{20\tau^7}  -  \frac{ 7( \lambda^2  -19 ) \alpha^2}{400  \tau^{10}} \right)J^3_- +
\frac{11\alpha^2}{400 \tau^{12}} J^5_-,
\cr
a_- &=& -\frac{ \alpha  }{  \bar{\tau}    } \left(  a^3_+ \vert_{traceless} + \frac{3\lambda^2 - 7}{20\tau^2} 
a_+  \right). 
\eea
Note that $a^3_+$ is originally traceless. We are keeping terms up to quadratic order in $\alpha$. Again keeping to the same notations as in the spin-3 case,
\bea
\Lambda^0 &=& ZT \left(  J_+ + \frac{1}{4T^2} J_- \right),\cr
\Lambda^1 &=& e^{3\rho} Z \left(  \frac{J^3_-}{20T^6} \right) -
e^{3\rho} \bar{Z} \left[  \left( J_+ + \frac{1}{4\tau^2} J_- \right)^3 + \frac{3\lambda^2 - 7}{20T^2} 
\left( J_+ + \frac{1}{4T^2} J_- \right) \right], \cr
\Lambda^2 
&=& e^{6\rho} \Bigg[
Z \left(  \frac{3(\lambda^2 - 9)(\lambda^2 - 4)}{400T^7} J_- -  \left(\frac{ 7 (\lambda^2 - 19  ) \alpha^2}{400  T^{9}} \right)J^3_- + \frac{11}{400T^{11}} J^5_-  \right)\cr
&-&  \bar{Z} \left[  \frac{3\lambda^2 - 7}{400T^9}J^3_- + \frac{1}{20Z^2T^9}\left( 
\left( \Lambda^0_\rho\right)^2 J^3_- +J^3_-\left( \Lambda^0_\rho\right)^2 + \left( \Lambda^0_\rho\right) J^3_- \left( \Lambda^0_\rho\right)\right)     \right] \Bigg].
\eea
The derivation proceeds as before and thus we will skip most details. 
After some straightforward algebra and invoking the techniques of Section \ref{sec:Prelim}, we find that the first-order correction can be written in the form
\be
\Phi^{(1)}(\rho,z,\bar{z}) = \Phi^{BTZ}(\rho,z,\bar{z})\frac{(\lambda + 3)(\lambda + 2)(\lambda + 1)}{\tau^3 ( \cos \frac{Z}{2} \cos \frac{\bar{Z}}{2} + 4T \bar{T} \sin \frac{Z}{2}  \sin \frac{\bar{Z}}{2}  )^3} \sum_{k=0}^3 \mathcal{S}^{(k)}(Z,\bar{Z}) (T\bar{T})^k
\ee
where the various terms read
\bea
\mathcal{S}^{(0)} (Z,\bar{Z}) &=& \frac{1}{40}\sin \frac{Z}{2} \cos^3 \frac{\bar{Z}}{2} \left( (Z-\bar{Z})(\cos Z -4) +3 
\sin Z \right), \cr
\mathcal{S}^{(1)} (Z,\bar{Z}) &=& \frac{1}{20}\sin \frac{\bar{Z}}{2} \cos^2 \frac{\bar{Z}}{2} \left( 3(Z-\bar{Z})(3\cos \frac{Z}{2} - \cos \frac{3Z}{2}) -27 \sin \frac{Z}{2} + 5 \sin \frac{3Z}{2}
\right), \cr
\mathcal{S}^{(2)} (Z,\bar{Z}) &=& -\frac{6}{5} \cos \frac{\bar{Z}}{2} \sin^2 \frac{\bar{Z}}{2} \sin \frac{Z}{2} 
\left( (Z-\bar{Z})(\cos Z + 2) - 3\sin Z    \right), \cr
\mathcal{S}^{(3)} (Z,\bar{Z}) &=& \frac{4}{15} \sin^3 \frac{\bar{Z}}{2} 
\left(  3(Z-\bar{Z}) (\cos \frac{3Z}{2} + 9 \cos \frac{Z}{2} ) - 27 \sin \frac{Z}{2} - 11 \sin \frac{3Z}{2}    \right).
\eea
In the infinite $\rho$ limit, only the term containing $\mathcal{S}^{(3)}$ contributes. The boundary two-point function reads
\bea
\label{spin4FirstOrder}
\lim_{\rho \rightarrow \infty}\Phi^{(1)} (\rho,z,\bar{z}) &=& \lim_{\rho \rightarrow \infty}\Phi^{BTZ}(\rho,z,\bar{z}) \times \frac{(\lambda + 3)(\lambda + 2)(\lambda +1) }{(16)(15)\sin^3 \frac{Z}{2}\tau^3} \Bigg(  3(Z-\bar{Z}) (\cos \frac{3Z}{2} + 9 \cos \frac{Z}{2} ) \cr
&&- 27 \sin \frac{Z}{2} - 11 \sin \frac{3Z}{2}    \Bigg). 
\eea
We will reproduce this expression shortly from the perspective of the dual CFT deformed by the appropriate higher-spin chemical potential. 

As mentioned earlier, we find that the second-order contribution to the propagator contains a large
number of terms for which we are unfortunately unable to simplify, even in the limit of infinite $\rho$.  
For all it is worth, in Appendix \ref{sec:AppA}, we express the bulk-boundary propagator in a closed form that is slightly more explicit than \eqref{3secondorder}. For the spin-4 black hole, analogous to the spin-3 black hole, we find that the bulk-boundary propagator can be cast in the form
\bea
\Phi^{(2)}(\rho,z,\bar{z}) &=& \Phi^{BTZ}(\rho,z,\bar{z}) \frac{u_f}{T^6 ( \cos \frac{Z}{2} \cos \frac{\bar{Z}}{2} + 4T \bar{T} \sin \frac{Z}{2}  \sin \frac{\bar{Z}}{2}  )^6 } \sum^6_{n=0} \mathcal{V}^{(n)} \left( Z, \bar{Z} \right) \left( T\bar{T} \right)^n, \cr
u_f &=& \frac{(\lambda +3)(\lambda + 2)(\lambda +1)}{6}.
\eea
In the infinite $\rho$ limit, only the term containing $\mathcal{V}^{(6)}$ contributes. In Appendix \ref{sec:AppB}, we will nonetheless furnish a dual CFT calculation (following the spin-3 calculation in \cite{ProbeCFT}) that by virtue of the holographic duality, gives a prediction for $\mathcal{V}^{(6)}$. Another approach to computing this boundary two-point function from the bulk directly might be to apply the methodology of \cite{resum} to this case. As mentioned earlier, this would rely on the validity of performing an analytic continuation from some integer to arbitrary $\lambda$.

\section{Boundary two-point functions from the dual CFT}
\label{sec:Boundary}
Finally, we proceed to examine the scalar propagator from the perspective of the putative dual CFT which 
has a $W_\infty [\lambda]$ symmetry and which is deformed by holomorphic higher-spin currents.
\be
\delta S_{CFT} = \frac{1}{\pi} \int d^2 z \,\,\mu (z, \bar{z}) U(z)
\ee
where $U(z)$ is some higher-spin current (we will take it to be a spin-4 current in this section). The gravity background is that of the BTZ black hole deformed by a chemical potential for the higher-spin charges so the CFT resides on its boundary torus parametrized by the complex coordinates
\be
(z,\bar{z}) \sim (z+2\pi \tau, \bar{z} + 2\pi \bar{\tau}).
\ee
The holomorphic chemical potential is defined by taking $\mu = \alpha/ \bar{\tau}$. The torus can be realized on the complex plane as an annulus with its boundaries identified. The deformed two-point function is then the torus correlation function 
\be
\text{Tr} \left( \bar{\phi}(z_1, \bar{z}_1) \phi(z_2, \bar{z}_2) e^{-\delta S_{CFT}} q^{L_0 -\frac{c}{24}} \bar{q}^{\bar{L}_0 - \frac{c}{24}} \right),
\ee
where $\phi$ is the scalar primary operator in the CFT with $h=\bar{h}=\frac{1}{2}(\lambda + 1)$. 
Switching to annulus coordinates $w=e^{iz}$, and invoking the usual transformation property of quasi-primary
fields, at first order we then have
\be
-\frac{\mu}{\pi} \int d^2 v \frac{v^3}{\bar{v}} w^h_1 \bar{w}^h_1 w^h_2 \bar{w}^h_2 
\text{Tr} \left(  U(v) \bar{\phi} (w_1, \bar{w}_1) \phi (w_2, \bar{w}_2) q^{L_0 - \frac{c}{24}}
\bar{q}^{\bar{L}_0 - \frac{c}{24}} \right).
\ee
We are interested in the regime of high temperature, and thus we perform a modular transformation to take $\tau \rightarrow -1/\tau$. In the high temperature, the leading contribution arises from the vacuum state 
and this yields (from now, we denote the boundary two-point function by $\Phi_B$) 
\be
\label{first}
\Phi^{(1)}_B =
-\frac{i\alpha}{2\pi}w^h_1 \bar{w}^h_1 w^h_2 \bar{w}^h_2 \tau^{2h+3}\bar{\tau}^{2h} \oint dv v^3 \text{ln} (v\bar{v}) \langle U(v) \bar{\phi}(w_1) \phi(w_2) \rangle_0,
\ee
where the subscript zero implies the correlation function on a two-sphere, and we have invoked Stokes' theorem and adopted a regularization scheme (see \cite{ProbeCFT,Dijk} ) where
\be
\label{regular}
\frac{1}{\bar{v}} = \bar{\partial} \text{ln} (v\bar{v} ).
\ee
The contour is taken along the holes cut around the insertion points $w_{1,2}$, although as explained
in the spin-3 calculation in \cite{ProbeCFT}, there are regular parts along the annulus's boundaries which could be shown to vanish with the use of \eqref{regular}. 
By virtue of the fact that $U,\bar{\phi},\phi$ are quasi-primary fields, the 3-point function
is fixed to read 
\be
\langle U(v) \bar{\phi}(w_1) \phi (w_2) \rangle = \frac{u_f w^4_{12}}{  w^{2h}_{12}  \bar{w}^{2h}_{12} (v-w_1)^4 (v-w_2)^4 }
\ee
where $u_f$ denotes the spin-4 eigenvalue \cite{SymMin} 
\be
u_f = \frac{(\lambda +1)(\lambda + 2)(\lambda +3)}{20}.
\ee
From \eqref{first}, we thus have 
\be
\Phi^{(1)}_B = -\frac{i\alpha}{2\pi}w^h_1 \bar{w}^h_1 w^h_2 \bar{w}^h_2 \tau^{2h+3}\bar{\tau}^{2h} \frac{u_f w^4_{12}}{  w^{2h}_{12}  \bar{w}^{2h}_{12}} \oint dv \frac{v^3 \text{ln} (v\bar{v}) }{(v-w_1)^4 (v-w_2)^4 }.
\ee
The contour integral has poles at $w_1$ and $w_2$ and can be easily performed to yield 
\be
\label{FirstOrder}
\Phi^{(1)}_B = \frac{\alpha \tau^3 \Phi^{BTZ}_B u_f}{12 \sin^3 \frac{\tau z}{2}}
\left[  \left(11 \sin \frac{3\tau z}{2} + 27 \sin \frac{\tau z}{2} \right) 
-3 (\tau z - \bar{\tau} \bar{z} ) \left( \cos \frac{3\tau z}{2} + 9 \cos \frac{\tau z}{2} \right) \right]
\ee
where we have taken $z_1 = z, z_2=0, w_1 = e^{i\tau z}, w_2 = 1$, following our convention for the
bulk-boundary propagator in the earlier sections. 

Incidentally, we can also derive the 3-point function by a slightly longer route which is relevant for computing the 4-point function later. 
We need the OPE between the spin-4 field and the scalar field in the large central charge limit. It reads
\be
\label{OPEUPhi}
U(v) \bar{\phi}(w_1) = \frac{(U_0 \bar{\phi})(w_1)}{(v-w_1)^4} + 
\frac{(U_{-1} \bar{\phi})(w_1)}{(v-w_1)^3} +
\frac{(U_{-2} \bar{\phi})(w_1)}{(v-w_1)^2} + 
\frac{(U_{-3} \bar{\phi})(w_1)}{(v-w_1)},
\ee
where 
\bea
U_0 \bar{\phi} &=& u_f \bar{\phi}, \cr
U_{-1} \bar{\phi} &=& \frac{1}{5}(\lambda +3 ) (\lambda +2) \partial \bar{\phi}, \cr
U_{-2} \bar{\phi} &=& \frac{1}{2}(\lambda + 3) \partial^2 \bar{\phi}, \cr
U_{-3} \bar{\phi} &=& \partial^3 \bar{\phi}.
\eea
These relations were worked out in Section 5.3 of \cite{Raju} by deriving the null vector structure up to level three in the 't Hooft limit of the minimal models. The contour integral can be straightforwardly computed around the poles of the integrand. From the poles at $w_1$, we obtain 
\bea
&&\oint dv \, v^3 \text{ln} (\bar{v} v ) \left( 
\frac{u_f}{(v-w_1)^4} + \frac{(\lambda +3)(\lambda + 2) \partial_{w_1}}{5(v-w_1)^3}
+ \frac{(\lambda +3) \partial^2_{w_1}}{2(v-w_1)^2} + \frac{\partial^3_{w_1}}{v-w_1} \right) \langle \bar{\phi}(w_1) \phi(w_2) \rangle_0 \cr
&=& \text{ln} (w_1 \bar{w}_1) \left[ w^3_1 \partial^3_{w_1} + \frac{3}{2}(\lambda +3) w^2_1 \partial^2_{w_1} + \frac{3}{5}(\lambda +3)(\lambda+2) w_1 \partial_{w_1} + u_f \right] \langle \bar{\phi}(w_1) \phi (w_2) \rangle  \cr
&&+ \left[ \frac{1}{2}(\lambda + 3) w^2_1 \partial^2_{w_1} + \frac{1}{2}(\lambda +3)(\lambda +2) w_1 \partial_{w_1} + \frac{11}{6}u_f \right] \langle \bar{\phi}(w_1) \phi (w_2) \rangle, 
\eea
where 
\be
\langle \bar{\phi}(w_1) \phi (w_2) \rangle= w^{-2h}_{12} \bar{w}^{-2h}_{12}.
\ee
The poles at $w_2$ contribute analogously, and putting the two terms together we obtain the expression \eqref{FirstOrder}. This is precisely the expression obtained via the bulk calculation in \eqref{spin4FirstOrder}.This entry of the higher-spin holographic dictionary anchors simply upon conformal invariance and choosing a suitably normalized higher-spin eigenvalue of the scalar operator in the dual CFT.

\section{Discussion}
\label{sec:Discussion}

In this paper, we have explored some aspects of three-dimensional higher-spin holography by studying scalar fluctuations in the background of higher-spin black holes, with the main novelty being an independent derivation of the bulk-boundary propagator using an infinite-dimensional matrix representation of $hs[\lambda]$ algebra from its construction as a quotient of the universal enveloping algebra of $sl(2)$. This evades the need in previous literature to perform an analytic continuation from some integer to $\lambda$, and furnishes a class of examples relevant for the study of the 't Hooft limit of higher-spin $AdS_3/CFT_2$ holography. The bulk-boundary propagators are computed for black hole solutions (in $hs[\lambda] \times hs[\lambda]$ Chern-Simons theory) deformed by spin-3 chemical potential up to second-order in the spin-3 potential. 
Taking the bulk-point to the boundary, we verify that the second-order correction of the boundary two-point function  matches that of the corresponding dual CFT's scalar correlator as computed in \cite{ProbeCFT}. At this order, such an identification of the entries of the holographic dictionary goes beyond constraints imposed by conformal invariance, and is sensitive to the form of the classical $W_\infty [\lambda]$ algebra involving spin-3, spin-4 fields,
the energy-momentum tensor and their descendents. 

We also applied this method to understand the scalar propagator for a black hole with spin-4 charge. 
The bulk-boundary propagator was computed to first order in the higher-spin chemical potential. On the CFT side, we generalize the calculation in \cite{ProbeCFT} to the case where the dual CFT with $\mathcal{W}_{\infty} [\lambda]$ is deformed by a holomorphic spin-4 operator. We match the correlators on both sides to first-order in the higher-spin potential.
Unfortunately, we encounter a huge number of terms in the bulk calculation of the second-order correction
and were not able to extract the boundary two-point function from them. Nonetheless, the method that we use allows us to express it in a closed form in terms of elementary integrals. We find that  the corresponding dual CFT computation is doable and we present the complete and explicit two-point function up to second-order in Appendix \ref{sec:AppB}. 

Overall, we hope that our work has shed light on the computational feasibility for similar computations, such as the calculation of $hs[\lambda]$-valued Wilson lines as recently discussed in \cite{GenHS} without relying on the analytic continuation from some integer to $\lambda$. Our work may also be useful for exploring fluctuations in other classical higher-spin backgrounds such as the conical spaces of \cite{Conical}, or black holes in 
higher-spin supergravity theories as discussed in \cite{SUGRA1,SUGRA2,SUGRA3}. Beyond computing boundary two-point functions, the technique presented here would be useful for studying Witten diagrams and the role of conformal blocks in higher-spin holography as recently advocated in \cite{Hijano,Besken}.

\section*{Acknowledgments}

It is a pleasure to thank Jan de Boer, Alejandra Castro, Ori Ganor, Petr Ho\v{r}ava, Juan Jottar, Per Kraus and Eric Perlmutter for various discussions on the theme of higher-spin gravity. 
I acknowledge support from the Foundation for Fundamental Research on Matter (FOM) which is part of the Netherlands Organization for Scientific Research as well as a research fellowship of NTU during the course of completion of this work. 

\appendix
\section{On second-order contribution to the bulk-boundary propagator in the spin-4 black hole background}
\label{sec:AppA}
In this section, we'll make \eqref{3secondorder} more explicit for the spin-4 black hole. Writing
\bea
\Lambda^{2} &=& \mathcal{F}^-(Z) J_- + \mathcal{F}^{---}(Z,\bar{Z}) J^3_- + \mathcal{F}^{abcde}(Z,\bar{Z}) J_a J_b J_c J_d J_e \cr
\Lambda^1 &=& \mathcal{G}^i (\bar{Z}) J_i + \mathcal{G}^{ijk}(Z,\bar{Z}) J_{ijk},\cr
\mathcal{B}^i(s,Z,\bar{Z}) &=& \mathcal{G}^j(\bar{Z}) \mathcal{C}_j^i (s,Z), \,\,\, 
\mathcal{B}^i(Z,\bar{Z}) = \int^1_0 ds\, \mathcal{B}^i (s,Z,\bar{Z}) \cr
\mathcal{B}^{ijk}(s,Z,\bar{Z}) &=& \mathcal{G}^{lmn}(Z,\bar{Z}) \mathcal{C}_l^i(s,Z)     \mathcal{C}_m^j(s,Z)
\mathcal{C}_n^k(s,Z) ,\,\,\, \mathcal{B}^{ijk} (Z,\bar{Z})=\int^1_0 ds\, \mathcal{B}^{ijk}(s,Z,\bar{Z}),\nonumber
\eea
the second-order correction to propagator is the sum of the contributions due to the first- and second-order corrections to the Chern-Simons connection which we denote as $\left[ e^{\bar{\Lambda}}  \Phi_{\Lambda^2} e^{-\Lambda^0} \right]_{11}$, $\left[ e^{\bar{\Lambda}} \Phi_{\Lambda^1}e^{-\Lambda^0} \right]_{11}$ respectively, with
\bea
\Phi_{\Lambda^2}&=&-2 \int^1_0 ds \, \mathcal{F}^-(Z) \mathcal{C}^i_- (s,Z) J_i + 
\mathcal{F}^{---}(Z,\bar{Z}) \mathcal{C}^i_- (s,Z) \mathcal{C}^j_- (s,Z) \mathcal{C}^k_- (s,Z) J_i J_j J_k  \cr
&+& \mathcal{F}^{abcde}(Z,\bar{Z}) \mathcal{C}^i_a (s,Z) \mathcal{C}^j_b (s,Z)
\mathcal{C}^k_c (s,Z) \mathcal{C}^l_d (s,Z) \mathcal{C}^m_e (s,Z) J_i J_j J_k J_l J_m,
\eea
\bea
\Phi_{\Lambda^1}&=& \int^1_0 ds\,\left[ \mathcal{B}^i (sZ,s\bar{Z}) \mathcal{B}^j (s,Z,\bar{Z}) + \mathcal{B}^i (s,Z,\bar{Z})\mathcal{B}^m (-sZ,-s\bar{Z}) \mathcal{C}_m^j(s,Z) \right]J_i J_j   \cr
&&+ \Bigg[ \mathcal{B}^{ijk}(sZ,s\bar{Z}) \mathcal{B}^l(s,Z,\bar{Z}) +
\mathcal{B}^{i}(sZ,s\bar{Z}) \mathcal{B}^{jkl}(s,Z,\bar{Z}) \cr
&&+\mathcal{B}^{pqr}((1-s)Z,(1-s)\bar{Z}) \mathcal{B}^i(s,Z,\bar{Z}) 
\mathcal{C}^j_p (s,Z) \mathcal{C}^k_q (s,Z) \mathcal{C}^l_r (s,Z) \cr
&&+ \mathcal{B}^{n}((1-s)Z,(1-s)\bar{Z})\mathcal{C}^i_n (s,Z) \mathcal{B}^{ijk}(s,Z,\bar{Z})
\Bigg] J_i J_j J_k J_l \cr
&&+\left[ \mathcal{B}^{ijk}(sZ,s\bar{Z}) \mathcal{B}^{lmn}(s,Z,\bar{Z}) 
+ \mathcal{B}^{pqr}((1-s)Z,(1-s)\bar{Z}) \mathcal{C}^l_p (s,Z) \mathcal{C}^m_q (s,Z) \mathcal{C}^n_r (s,Z) 
\mathcal{B}^{ijk}(s,Z,\bar{Z}) \right] \cr
&&\qquad \qquad \qquad \qquad \times J_i J_j J_k J_l J_m J_n. 
\eea
Although the integrals are elementary, we were unfortunately unable to simplify the huge number of terms present. For the purpose of obtaining the boundary two-point function, another possibility would be to study if the method of \cite{resum} enables a more manageable computation for this higher-spin background. Nonetheless, we now proceed to furnish a dual CFT calculation that should match this bulk correction
in the infinite $\rho$ limit.

\section{Second-order correction from four-point functions in the dual CFT}
\label{sec:AppB}

The second-order term involves a four-point function and it reads
\be
\Phi^{(2)}_B=
-\frac{\alpha^2}{8\pi^2}w^h_1 \bar{w}^h_1 w^h_2 \bar{w}^h_2 
\tau^{2h+6} \bar{\tau}^{2h} \oint dv_1 \oint dv_2 \text{ln}(v_1 \bar{v}_1 ) \text{ln}(v_2 \bar{v}_2) v^3_1 v^3_2 \langle U(v_1) U(v_2) \bar{\phi} (w_1) \phi (w_2) \rangle.
\ee
Relative to the first-order calculation, the second-order one is more involved and relies in particular on the OPE between two spin-4 fields which reads\footnote{We essentially deduce this OPE from Section 3.3 of \cite{SymMin} and \cite{Kwang}. The result in eqn.26 of \cite{Zhang} was also useful.} 
\bea
U (v_1) U(v_2) &=& \frac{6Y(v_2)}{(v_1 - v_2)^2} + \frac{3\partial Y (v_2)}{(v_1 - v_2)} \cr
&&+ 36 n_{44} \left[ \frac{U(v_2)}{(v_1-v_2)^4} + \frac{\partial U(v_2)}{2v^3_{12}} + \frac{5}{36 v^2_{12}} \partial^2 U(v_2) + \frac{1}{36v_{12}} \partial^3 U(v_2) \right] \cr
&&-\frac{14 N_4}{3} \left[    
\frac{2T(v_2)}{v^6_{12}} + \frac{\partial T(v_2)}{v^5_{12}} + \frac{3}{10} \frac{\partial^2 T(v_2)}{v^4_{12}} + \frac{1}{15} \frac{\partial^3 T (v_2)}{v^3_{12}} 
+\frac{1}{84} \frac{\partial^4 T (v_2)}{v^2_{12}} + \frac{1}{560} \frac{\partial^5 T(v_2)}{v_{12}} 
\right]  \cr
\label{w4algebra}
&&+ \ldots 
\eea
where $Y,U,T$ are spin-6, 4,2 fields and we have omitted terms which can be ignored in the large $c$ limit. They include composite fields and quantum corrections to the classical $\mathcal{W}$-algebra. The constants read 
\be
n_{44} = \frac{8}{15}\sigma^2 (\lambda^2 - 19), N_4 = -\frac{384}{85}\sigma^4 (\lambda^2 - 4)(\lambda^2 -9)
\ee
where $\sigma$ is some arbitrary constant which we can fix by comparison to the bulk gravity result. The four-point function can be computed using the above OPE as well as that between the spin-4 field and the scalar field, i.e. 
\be
\bcontraction{}{U}{}{U} UU\bar{\phi}\phi + 
\bcontraction{}{U}{U}{\bar{\phi}} UU\bar{\phi}\phi +
\bcontraction{}{U}{U\bar{\phi}}{\phi} UU\bar{\phi}\phi.
\ee
Consider the contribution coming from the $UU$ OPE, and in particular the two terms involving the spin-6 field $Y$. Up to other multiplicative terms, we need to compute the integral
\be
\int \int \frac{d^2 v_2}{\bar{v}_2} \frac{d^2 v_1}{\bar{v}_1} v^3_1 v^3_2 \left(  \frac{6}{v^2_{12}} + \frac{3\partial_{v_2}}{v_{12}} \right) \langle Y(v_2) \bar{\phi}(w_1) \phi (w_2) \rangle_0.
\ee
We perform integration by parts for the second term to remove the derivative $\partial_{v_2}$ and then for both terms integrate over $v_1$ first, using Stokes' theorem after replacing $1/\bar{v}_1$ with 
$\text{ln}(v_1 \bar{v}_1)$. This gives us
\be
\int \frac{d^2 v_2}{\bar{v}_2}3v^5_2 \langle Y(v_2) \bar{\phi}(w_1) \phi (w_2) \rangle.
\ee
The three-point function $\langle Y(v_2) \bar{\phi}(w_1) \phi (w_2) \rangle$ can be obtained by using the fact that they are all quasi-primary fields. It reads 
\be
\langle Y(v_2) \bar{\phi}(w_1) \phi (w_2) \rangle = \frac{u_6}{w^{2h}_{12} \bar{w}^{2h}_{12}} \left(   \frac{w_{12}}{(v_2-w_1)(v_2-w_2)}   \right)^6,
\ee
where $u_6$ is the spin-6 eigenvalue which we determine\footnote{We follow Section 4 of \cite{SymMin}, computing the spin-6 eigenvalue by expressing $V^6_0$ in terms of a polynomial in $J_0$ and other terms which annihilate the scalar field. } 
to be
\be
u_6 = -\frac{1}{126}(\lambda + 5)(\lambda + 4)(\lambda + 3)(\lambda + 2)(\lambda + 1).
\ee
The other terms arising from the spin-4 and spin-2 fields can be handled similarly, and together with the other factors, we find that the second-order correction reads
\bea
\label{secondorder}
\Phi^{(2)}_{UU} &=& -\frac{\alpha^2}{8\pi^2} \Phi^{BTZ} \tau^6(2\pi i) \times \cr
&&\Bigg( 3 u_6 w^6_{12} \oint dv_2 \frac{v^5_2 \text{ln}(\bar{v}_2 v_2)}{\left[  (v_2 - w_1)(v_2 -w_1) \right]^6}
-7 n_{44} u_4 w^4_{12} \oint dv_2 \frac{v^3_2 \text{ln}(v_2 \bar{v}_2)}{\left[ (v_2 - w_1)(v_2 - w_2)  \right]^4} \cr
&&- \frac{3h}{10} w^2_{12} \oint dv_2 \frac{v_2 \text{ln}(v_2 \bar{v}_2)}{(v_2-w_1)^2 (v_2 - w_2)^2} \Bigg),
\eea
where 
\be
\Phi^{BTZ} = \frac{(w_1 \bar{w}_1 w_2  \bar{w}_2 )^h \tau^{2h} \bar{\tau}^{2h}  }{w^{2h}_{12} \bar{w}^{2h}_{12}},
\ee
and each of the line integrals can be computed to yield
\be
\oint dv_2 \frac{v_2 \text{ln}(v_2 \bar{v}_2)}{(v_2-w_1)^2 (v_2 - w_2)^2} = -2\pi i \left[ \frac{2}{w^2_{12}} + \frac{(w_1 + w_2)}{w^3_{12}} \text{ln}\left( \frac{w_2 \bar{w}_2}{w_1 \bar{w}_1} \right) \right]
\ee
\bea
\oint dv_2 \frac{v^3_2 \text{ln}(v_2 \bar{v}_2)}{\left[ (v_2 - w_1)(v_2 - w_2)  \right]^4} &=& \frac{-2\pi i}{6w^7_{12}} \Bigg[ 22(w^3_1 - w^3_2) + 54 w_1 w_2 w_{12} \cr
&+& 6 \text{ln} \left( \frac{w_2 \bar{w}_2}{w_1 \bar{w}_1} \right) (w^3_1 + w^3_2 + 9w_1 w_2 (w_1 + w_2) ) \Bigg] \cr
 \oint dv_2 \frac{v^5_2 \text{ln}(\bar{v}_2 v_2)}{\left[  (v_2 - w_1)(v_2 -w_1) \right]^6} &=& 
\frac{-2\pi i}{5!w^{11}_{12}} \Bigg[ 4 \left(137(w^5_1 - w^5_2) + 1625 w_1 w_2 (w^3_1 - w^3_2) + 2000 w^2_1 w^2_2 (w_1 - w_2) \right) + \cr
&+& 120 \text{ln} \left( \frac{w_2 \bar{w}_2}{w_1 \bar{w}_1} \right) \left(  w^5_1 + w^5_2 + 25 w_1 w_2 (w^3_1 + w^3_2) + 100 w^2_1 w^2_2 (w_1 + w_2) \right) \Bigg]. \nonumber \\
\eea
Just as in the first-order computation , we take $w_1 = e^{iz}, w_2 = 1$, and we have from the Wick contraction of the two spin-4 fields
\bea
\label{UUsecond}
\Phi^{(2)}_{UU} &=& \frac{\alpha^2}{2}\Phi^{BTZ} \tau^6 \times 
\Bigg[ h_1(\lambda) (2 - \cot \frac{z \tau}{2} (z \tau - \bar{z} \bar{\tau}) ) + \frac{h_2 (\lambda)}{\sin^5\frac{z \tau}{2}} 
\Bigg[ (137 \sin\frac{5z \tau}{2} + 1625 \sin \frac{3z \tau}{2} + 2000 \sin \frac{z \tau}{2} ) \cr
&&-30(\cos \frac{5z \tau}{2} + 25 \cos \frac{3z \tau}{2} + 100 \cos \frac{z \tau}{2} )(z \tau-\bar{z} \bar{\tau}) \Bigg] + \frac{h_3(\lambda)}{\sin^3 \frac{z \tau}{2}} \Bigg( (22 \sin \frac{3z \tau}{2} + 54 \sin \frac{z \tau}{2} ) \cr
&&-6(z \tau-\bar{z} \bar{\tau})(\cos \frac{3z \tau}{2} + 9 \cos \frac{z \tau}{2}) \Bigg) + \ldots
\eea
where the $\lambda-$dependent coefficients read
\bea
h_1 (\lambda) &=& \frac{288}{425} (\lambda +1)(\lambda^2 - 4)(\lambda^2 - 9) \sigma^4, \cr
h_2 (\lambda) &=& - \frac{1}{20160}(\lambda+5)(\lambda+4)(\lambda+3)(\lambda+2)(\lambda+1), \cr
h_3 (\lambda) &=& \frac{7}{900} (\lambda^2 - 19)(\lambda+3)(\lambda+2)(\lambda+1)\sigma^2
\eea
with $\sigma$ being an arbitrary constants. The various constants descend from the precise form 
of the classical $\mathcal{W}_\infty [\lambda]$ algebra involving the relevant higher-spin fields. We are also taking the large $c$ limit. 
We are now left with evaluating the Wick contractions 
\be
\bcontraction{}{U}{U}{\bar{\phi}} UU\bar{\phi}\phi +
\bcontraction{}{U}{U\bar{\phi}}{\phi} UU\bar{\phi} \phi .
\ee
They can be computed straightforwardly via the OPE \eqref{OPEUPhi}. Define 
\bea
\mathcal{D} (w_1) &=&
 \text{ln} (w_1 \bar{w}_1) \left[ w^3_1 \partial^3_{w_1} + \frac{3}{2}(\lambda +3) w^2_1 \partial^2_{w_1} + \frac{3}{5}(\lambda +3)(\lambda+2) w_1 \partial_{w_1} + u_f \right] \langle \bar{\phi} \phi \rangle  \cr
&&+ \left[ \frac{1}{2}(\lambda + 3) w^2_1 \partial^2_{w_1} + \frac{1}{2}(\lambda +3)(\lambda +2) w_1 \partial_{w_1} + \frac{11}{6}u_f \right].
\eea
The contribution due to the two contractions is then given by
\be
-\frac{\alpha^2}{2}(w_1 \bar{w}_1 w_2 \bar{w}_2)^h \tau^{2h+6} \bar{\tau}^{2h} \left[ \mathcal{D}(w_1) + \mathcal{D}(w_2) \right]^2 \frac{1}{w^{2h} \bar{w}^{2h} } .
\ee
Taking $w_1= e^{iz}, w_2=1$, it reads after simplification
\bea
\label{Uphisecond}
\Phi^{(2)}_{U\Phi} &=& \frac{\alpha^2}{2}\Phi^{BTz\tau} \tau^6 
\left( \frac{u_4}{11520 \sin^6\frac{z\tau}{2}} \right) \Bigg[ 
4(\lambda +3) \sin \frac{z\tau}{2} \Bigg( \sin \frac{z\tau}{2} (2f_1(\lambda) \cos 2z\tau + 8 f_2(\lambda) \cos 2z\tau + 6 f_3(\lambda) ) \cr
&&- 3 \cos \frac{z\tau}{2} ( 22 f_5(\lambda) \cos 2z\tau + 36 f_6 (\lambda) \cos z\tau + 2 f_7 (\lambda))(z\tau-\bar{z} \bar{\tau}) \Bigg)  + 6 \sin z\tau \bigg( 2g_1(\lambda) \cos 2z\tau  \cr
&&+ 36 g_2(\lambda) \cos z\tau + g_4 (\lambda) \bigg)(z\tau-\bar{z} \bar{\tau})-9 ( 2g_5 (\lambda) \cos 3z\tau + 36 g_6 (\lambda) \cos 2z\tau + 18 g_7 (\lambda) \cos z\tau \cr
&&+ 4g_8(\lambda) ) (z\tau-\bar{z} \bar{\tau})^2 \Bigg],
\eea
where the $\lambda-$dependent coefficients read
\bea
&&g_1 (\lambda) = g_3 (\lambda) = 132 + 211 \lambda + 102 \lambda^2 + 11\lambda^3, \cr
&&g_2 (\lambda) = 564 + 457\lambda + 104 \lambda^2 + 7 \lambda^3, \cr
&&g_4(\lambda) = 51432 + 27526\lambda + 5052 \lambda^2 + 326 \lambda^3, \cr
&&g_5( \lambda) = 6 + 11 \lambda + 6 \lambda^2 + \lambda^3, \cr
&&g_6 (\lambda) = 54 + 51 \lambda + 14 \lambda^2 + \lambda^3, \cr
&&g_7 (\lambda) = 1242 + 801 \lambda + 162 \lambda^2 + 11 \lambda^3,\cr
&& g_8 (\lambda) = 5922 + 3331 \lambda + 642 \lambda^2 + 41 \lambda^3 \cr
&& f_1 (\lambda) = 422 + 543 \lambda + 121 \lambda^2,\cr
&& f_2 (\lambda) = 3748 + 1977 \lambda + 209 \lambda^2,\cr
&& f_3(\lambda) = 6862 + 2583 \lambda + 281 \lambda^2,\cr
&& f_5(\lambda) = (\lambda+1)(\lambda+2),\cr
&& f_6(\lambda) = 94 + 61 \lambda + 7 \lambda^2,\cr
&& f_7(\lambda) = 4286 + 1569\lambda + 163 \lambda^2.
\eea
The overall second-order correction is then the sum of \eqref{UUsecond} and \eqref{Uphisecond}, i.e.
\be
\Phi^{(2)} = \Phi^{(2)}_{UU} + \Phi^{(2)}_{U\Phi}.
\ee

\end{document}